\pdfoutput=1
\documentclass[conference,10pt]{IEEEtran}
\ifCLASSINFOpdf
   \usepackage[pdftex]{graphicx}
   \graphicspath{{./}{./img/}}
   \DeclareGraphicsExtensions{.pdf,.jpeg,.jpg,.png,.eps,.ps}
\else
  \usepackage[dvips]{graphicx}
  \graphicspath{{./}{./img/}{./eps/}}
  \DeclareGraphicsExtensions{.pdf,.jpeg,.jpg,.png,.eps,.ps}
\fi
\usepackage{cite}
\usepackage[cmex10]{amsmath}
\usepackage{algorithmic}
\usepackage{array}
\usepackage{mdwmath}
\usepackage{mdwtab}
\usepackage{eqparbox}
\usepackage[font=footnotesize]{subfig}
\usepackage{stfloats}
\usepackage{url}

\usepackage[lined, ruled, linesnumbered,noend]{algorithm2e}
\usepackage[bookmarks=false]{hyperref}
\usepackage{latexsym}
\usepackage{amsfonts,amssymb,amsthm}
\usepackage{xcolor}

\usepackage{enumitem}
\usepackage{pbox}
\usepackage{mathtools}
\usepackage{float}
\usepackage{pgffor}
\usepackage{bm}		%
\usepackage{tabularx}	%
\usepackage{framed}	
\usepackage{cleveref}	%
\usepackage{pifont}		%
\usepackage[top=0.763in, bottom=0.68in, left=0.68in, right=0.6in]{geometry}
\IEEEoverridecommandlockouts

\IEEEaftertitletext{\vspace{0.3mm}}

\newtheorem{theorem}{\bf Theorem}		%
\newcounter{appdx}
\newcommand{\kETAL}     {{\em et~al.}}		%

\newcommand{\smallfont}  {}

\definecolor{dgreen}{rgb}{0,0.655,0.149}

\definecolor{dgreenn}{HTML}{00A64F}

\newcommand{\redc}[1]{{\color{red}}}

\SetKw{Continue}{continue}
\SetKw{Break}{break}

\newcommand{\myendbox}{\hfill \qed}

\makeatletter
\newcommand{\nosemic}{\renewcommand{\@endalgocfline}{\relax}}%
\newcommand{\dosemic}{\renewcommand{\@endalgocfline}{\algocf@endline}}%
\let\oldnl\nl%
\newcommand{\nonl}{\renewcommand{\nl}{\let\nl\oldnl}}%
\makeatother

\SetKwInOut{Init}{Initialize}
\SetKwInOut{OptIn}{Optional input}

\begin{document}


\title{QuESat: Satellite-Assisted Quantum Internet for Global-Scale Entanglement Distribution
\author{Huayue~Gu,~Ruozhou~Yu,~Zhouyu~Li,~Xiaojian~Wang,~Guoliang~Xue}
\IEEEcompsocitemizethanks{\IEEEcompsocthanksitem Gu, Yu, Li and Wang (\{hgu5, ryu5, zli85, xwang244\}@ncsu.edu) are with NC State University, Raleigh, NC 27606, USA. Xue (xue@asu.edu) is with Arizona State University, Tempe, AZ 85281, USA. 
Gu, Yu, Li, and Wang were supported in part by NSF grants 2045539 and 2350152. 
The research of Guoliang Xue was supported in part by the U.S. Department of Energy, Office of
Science, Advanced Scientific Computing Research (ASCR) program under contract number ERKJ432, as part of the PiQSci quantum networking project, and by NSF grant 2007083.
The information reported herein does not reflect the position or the policy of the funding agencies.

This paper has been accepted for publication at IEEE INFOCOM 2025.
}
}

\maketitle


\IEEEpeerreviewmaketitle

\begin{abstract}
Entanglement distribution across remote distances is critical for many quantum applications. 
Currently, the de facto approach for remote entanglement distribution relies on optical fiber for on-the-ground entanglement distribution.
However, the fiber-based approach is incapable of global-scale entanglement distribution due to intrinsic limitations. 
This paper investigates a new hybrid ground-satellite quantum network architecture (QuESat) for global-scale entanglement distribution, integrating an on-the-ground fiber network with a global-scale passive optical network built with low-Earth-orbit satellites.
The satellite network provides dynamic construction of photon lightpaths based on near-vacuum beam guides constructed via adjustable arrays of lenses, forwarding photons from one ground station to another with very high efficiency over long distances compared to using fiber. 
To assess the feasibility and effectiveness of QuESat for global communication, we formulate lightpath provisioning and entanglement distribution problems, considering the orbital dynamics of satellites and the time-varying entanglement demands from ground users.
A two-stage algorithm is developed to dynamically configure the beam guides and distribute entanglements, respectively.
The algorithm combines randomized and deterministic rounding for lightpath provisioning to enable global connectivity, with optimal entanglement swapping for distributing entanglements to meet users' demands.
By developing a ground-satellite quantum network simulator, QuESat achieves multi-fold improvements compared to repeater networks.

\end{abstract}




\ifCLASSOPTIONcompsoc
\IEEEraisesectionheading{\section{Introduction}\label{sec:intro}}
\else
\section{Introduction}
\label{sec:intro}
\fi
\noindent 
Entanglement distribution over remote distances is essential to many applications such as quantum key distribution (QKD)~\cite{bennett2020quantum}, distributed quantum computing (DQC)~\cite{cacciapuoti2019quantum,cicconetti2022resource}, and quantum internet-of-things~\cite{chen2022ddka}.
Current approaches mainly rely on optical fiber as light guides for entangled photons due to its flexibility, relatively low cost of deployment (compared to alternative approaches), and existing large-scale deployment.
However, exponential photon loss in fiber hinders single photon transmission as distances increase, limiting the practical range of such networks to a few hundred kilometers~\cite{photon-loss, nitish_satellite_2022}.

To overcome this limitation, the first-generation quantum networks aim to utilize quantum repeaters to enable long-distance entanglement distribution~\cite{Dur1999}.
A quantum repeater concatenates short-distance entanglements to generate longer-distance ones via entanglement swapping, and performs optional entanglement distillation to improve fidelity, with the help of quantum memories and measurement devices~\cite{pouryousef2022quantum,zhao2024Routing,zeng2022multi}.
While promising quantum repeaters have two limiting factors. 
First, their effectiveness is limited by the current low efficiency of memory and measurement devices, which degrades exponentially as more repeaters are needed to establish longer-distance entanglements. 
Second, quantum memory and measurement devices are particularly expensive and energy-hungry (including the need for cryostat for several candidate memory technologies), making them particularly costly and challenging to deploy and operate.

\begin{figure}[t]
\centering
\includegraphics[width=0.48\textwidth]{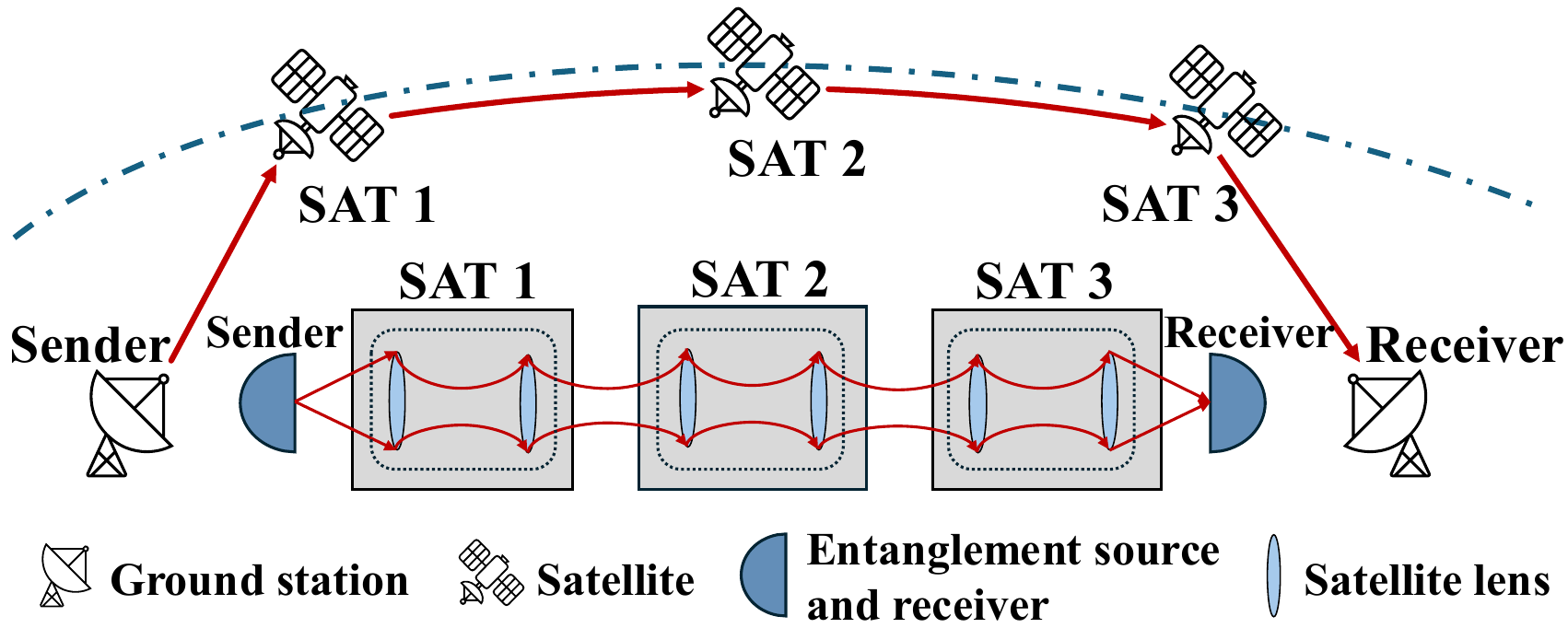}
\caption{\smallfont Architecture of satellite-assisted quantum internet.}
\label{fig:architecture}
\end{figure}
This paper investigates and proposes a new architecture for achieving global-scale entanglement distribution with 
ultra high efficiency. 
The architecture features, in addition to the ground fiber network, a satellite-assisted \emph{passive optical network} for directly transmitting entangled photons between arbitrary ground stations.
As shown in Fig.~\ref{fig:architecture}, each satellite is equipped with reflecting or refracting lenses, such that a sequence of lenses forms a free-space lightpath directly from the sender to the receiver for photon transmission.
To distribute an entanglement, the entanglement source (assumed to be at a ground station) generates entangled photon pairs, and sends one photon in each pair along the pre-configured lightpath directly to the receiver (another ground station).
Each satellite along the path simply reflects or refracts the photon stream without performing any operation, thereby removing the reliance on any expensive and energy-hungry memory or measurement device in space.
Furthermore, compared to fiber, such a free-space lightpath can have orders of magnitude lower loss (mainly due to reflection/refraction) at long distances, for instance, less than $10^{-3}$ to $10^{-4}$dB/km compared to 0.2dB/km for fiber~\cite{yin2017satellite,goswami2023satellite,huang2024vacuum}.
With a constellation of cost-efficient low-Earth-orbit (LEO) satellites covering the planet, and short-distance ground fiber connecting ground stations to nearby users, this architecture can provide highly-efficient and low-cost entanglement distribution over very long distances with currently available and near-term quantum technologies.

Compared to a repeater network based on swapping and/or distillation, the satellite passive optical network relies on pure mechanical operations---aligning the lenses for properly reflecting or refracting to the next hop/destination.
This prompts a new design problem of provisioning lightpaths to satisfy end-to-end entanglement needs.
Entanglements distributed over these lightpaths and/or via ground fiber can further be swapped by ground repeaters (e.g., co-located at the ground stations) to build entanglements between users or other ground stations who do not have an available lightpath.
The combined problem of lightpath provisioning and entanglement distribution becomes non-trivial and has no counterpart in either classical optical networks or quantum repeater networks alone.
To this end, we provide a formulation of this problem and then develop a two-stage algorithm to dynamically reconfigure satellite lightpaths based on satellite orbital dynamics, as well as to schedule swapping operations at repeaters to satisfy users' entanglement demands.
We further develop a hybrid ground-satellite quantum network simulator based on known satellite orbits.
Utilizing our algorithm and simulator, and combining existing simulation and experimental data, we evaluate the performance of the proposed hybrid architecture, and demonstrate that it achieves multi-fold improvements in entanglement rate and users' demand satisfaction compared to a fiber-based repeater network.

Our main contributions are as follows:

\begin{itemize}
    \item We propose a novel quantum network architecture that distributes highly efficient entanglements at a global scale through a combination of a ground repeater network and a satellite-assisted passive optical network without memories or measurement devices. The satellite network is implementable with current/near-term technologies as demonstrated by recent breakthroughs~\cite{goswami2023satellite,huang2024vacuum}.
    \item We formulate the lightpath provisioning and entanglement distribution problems in this hybrid ground-satellite architecture and develop a two-stage algorithm considering both the orbital dynamics of satellites and the dynamic demand changes of ground users.
    \item We develop a ground-satellite quantum network simulator, and use real satellite traces to demonstrate that the hybrid architecture can achieve multi-fold improvement in entanglement rate and demand satisfaction compared to ground repeater networks only with fiber.
\end{itemize}

\noindent\textbf{Organization.}
\S\ref{sec:rw} introduces background and related work.
\S\ref{sec:model} shows preliminaries for entanglement distribution in a satellite-assisted quantum network.
\S\ref{section:ps} presents the network model and problem formulation.
\S\ref{section:alg} proposes the two-stage algorithm design for entanglement distribution in QuESat.
\S\ref{sec:results} presents evaluation results.
\S\ref{sec:conclusions} is the conclusion.



\section{Background and Related Work}
\label{sec:rw}
\noindent
Since early demonstration in ideal lab conditions~\cite{peev2009secoqc,sasaki2011field,dahlberg2019link}, subsequent work has started exploring the practical-scale entanglement distribution networks for quantum applications.

\noindent\textbf{Fiber-based quantum repeater network:}
The first stage of research on entanglement distribution networks focuses on a fiber-based architecture~\cite{pirandola2019end}.
Short-distance nodes are connected by single-mode optical fiber to transmit entangled photons directly.
However, such transmission suffers from exponential loss of photons as the distance grows because of the attenuation loss in single-mode fiber.
To compensate for the excessive fiber loss, a repeater network, which can concatenate short-distance entanglements into longer-distance ones via swapping of repeaters connected by ground fiber, has become a main thrust~\cite{pant2019routing,photon-loss,mao2023qubit}.
Recent works have studied how to measure and profile a repeater path~\cite{liu2024quantum}, how to route entanglement traffic through a network of repeaters~\cite{pouryousef2022quantum,zhao2023scheduling,zhao2024Routing}, how to schedule entanglement swapping and distillation to maximize entanglement rate and limit decoherence~\cite{fendi}, and so on.
Nevertheless, the repeater network is intrinsically limited by several factors: its reliance on expensive devices, high energy consumption, high cost for production and large-scale deployment, all of which are key barriers to constructing a global quantum network with pure repeater technologies.

\noindent\textbf{Non-terrestrial entanglement distribution network:}
A promising approach for achieving global-scale entanglement distribution involves the use of free-space optics enabled by non-terrestrial nodes, including satellites or unmanned aerial vehicles (UAVs)~\cite{sen2023quantum_satellite,liu2021optical}.
The use of non-terrestrial nodes can circumvent the non-line-of-sight issue faced by ground transmission that warrants the use of fiber, thus significantly increasing flexibility while reducing photon loss.
Satellite-based entanglement distribution is specifically a promising alternative to the ground repeater network for global-scale entanglement distribution~\cite{yin2017satellite, yin2017satellite_2}.
Because of the restricted energy budget and payload constraint of satellites, existing works mainly utilize each individual satellite as an entanglement source that can send entangled pairs to two remote ground stations both within its line-of-sight~\cite{nitish_satellite_2022,wei2024optimal_sat,QCNC_sat}.
This naive architecture is, however, limited since a satellite's line-of-sight area cannot cover two locations, and no entanglement can be distributed.
It also significantly limits the number of users that each satellite can serve, as the satellite can only distribute entangled photons to a fixed pair of ground locations at a time.
Chang~\kETAL{}~\cite{chang2023entanglement_sat} proposed establishing multi-hop entanglement swapping paths with inter-satellite links by regarding satellites as quantum repeaters, which is not near-term feasible with limited satellite resources and the excessive overhead of repeater devices.

\noindent\textbf{Direct quantum communication with passive light guides:}
Since entangled photons are still a state of light, the classical idea of entanglement distribution via passive optical elements has recently been explored~\cite{goswami2023satellite, huang2024vacuum}.
The Vacuum Beam Guide (VBG)~\cite{huang2024vacuum} is a newly proposed ground-based quantum channel solution, which directly transmits photons through a sequence of kilometer-apart, aligned refraction lenses insulated within a vacuum chamber tube.
Eliminating fiber and other absorption loss, VBG can achieve as low as $10^{-4}$dB/km attenuation loss, 2-3 orders of magnitude lower than ground fiber.
Nevertheless, the downside of VBG includes its excessive ground deployment cost (over $10\times$ than fiber), and the anticipated high maintenance cost for the perfect alignment of lenses. 
In space, a similar idea has been proposed in~\cite{goswami2023satellite}, where a sequence of reflection mirrors mounted on LEO satellites can form a photon lightpath directly from one ground station to another and incurs only minimal reflection loss at each mirror plus absorption loss in the downlink and uplink through the Earth atmosphere.
The overall loss is expected to be as low as $30$dB at a $20,000$km ground distance, again two orders of magnitude lower than ground fiber.
In addition, these mirrors are extremely cheap compared to complex quantum devices such as memory or photon detectors, and do not consume any energy except for mechanical alignment to form the lightpaths.
These benefits make them the perfect candidate for a satellite-based, global-scale quantum communication infrastructure.



\section{Entanglement Distribution in a Ground-Satellite Network}
\label{sec:model}
\begin{table}
\caption{Notation Table}
\label{notation}
\begin{tabular}{p{1.9cm}p{6.2cm}}
\hline
Parameters & Description\\
\hline
$G=(V, E)$ & the quantum network with nodes $V$ and links $E$\\
$G^{\sf g} = (V^{\sf g}, E^{\sf g})
$ & the ground repeater network with ground stations $V^{\sf g}$ and fiber links $E^{\sf g}$\\
$G^{\sf s} = (V^{\sf s}, E^{\sf s})
$ & the satellite network with satellites $V^{\sf s}$ in orbits\\
$c_e^{\sf g}, q_e^{\sf g}, q_v^{\sf g}$ & fiber link capacity, ebit generation success probability \& swapping success probability \\
$c_v^{\sf s}, q_v^{\sf s}$ & satellite capacity \& transmission success probability \\
$q_e^{\sf gs}$ & photon survival probability in each GSL\\
$p, q(p), \alpha$ & the lightpath, path probability \& path capacity\\
$uv = \{  u, v\}$ & an unordered pair of nodes $m, n \in V^{\sf g} \cup V^{\sf s}$\\
$i, \mathcal{U}$ & the commodity \& the commodity set\\
$z^t_i, Z_i$ & demand at time $t$ \& demand set of commodity $i$\\
$s_i,d_i$ & commodity $i$'s source and destination ground stations  \\
$\bar s^t_i,\bar d^t_i$ & the satellites that
$s_i$ and $d_i$ are connected to at time $t$ \\
$m^t_i$ & the satellite topology change index at time t \\
\hline
Variables & Description\\
\hline
$x_i^{m^t_i}(uv)$ & integer variables to indicate if a lightpath for commodity $i$ utilizes an ISL $uv$ at time $t$\\
$f_i^{m^t_i}(u,v)$ & flow variables for commodity $i$ carried on an ISL $uv$\\
$\eta_i^{m^t_i}$ & total flow for lightpaths of commodity $i$\\
$y^{mk}_{mn}$ &  swapping rate from ${mk}$-ebits to ${mn}$-ebits.\\
$g_{mn}$ & elementary ebit generation ratio of $mn$ \\
$\zeta_i$ & total expected EDR for commodity $i$\\
\hline
\end{tabular}
\end{table}
\noindent
In this paper, we consider an entanglement distribution network whose goal is to distribute maximally entangled bipartite states (aka, \emph{Bell pairs} or \emph{ebits}) between remote ground stations. 
We use $|\Psi^-\rangle = \frac{1}{\sqrt{2}}( |01\rangle - |10\rangle )$ as our target state, although any of the four Bell pairs can be treated as the same during distribution.
We restrict our attention to bipartite entangled states since any multipartite entangled state (such as GHZ or W states) can be constructed using bipartite states distributed among multiple parties~\cite{hein2005entanglement}.
Below, we introduce common operations and unique characteristics between ground and satellite for quantum communication.
Notations are summarized in Table~\ref{notation}.

\subsection{Entanglement Generation and Manipulation}
\noindent

\noindent\textbf{Entanglement generation:} An entangled photon pair is generated by a physical process at the entanglement source, such as spontaneous parametric down-conversion
(SPDC)~\cite{spdc}, four-wave mixing~\cite{four-wave} or nitrogen-vacancy (NV) center~\cite{nv_1_46s}.
Each photon of a pair can be transmitted to other nodes via optical fibers, free space, or other media for further operations.

\noindent\textbf{Photon transmission:} Entangled photons transmitted through a quantum channel can get lost because of absorption, channel attenuation and other factors~\cite{photon-loss}. 
For example, the loss of photons increases exponentially with increasing distances when using optical fibers, typically in the order of $0.1$dB/km~\cite{dahlberg2019link}.
If a pair of photons both survive the transmission process (or if one is kept in a quantum memory while the other survives transmission), they become a remote entanglement between two nodes, which we call an \emph{elementary ebit}.
Considering the transmission loss, we view the remote entanglement generation over a quantum channel as a \emph{probabilistic process}, with the success probability denoted by $q_{{\sf link}} \in (0, 1]$.

In practical scenarios, a protocol is usually employed to try to deterministically generate remote ebits over a low-efficiency source and/or a high-loss channel.
One commonly adopted protocol is to attempt ebit generation for $N_{\sf attp}$ times during a unit time slot and store only one ebit if at least one is successfully generated.
We factor such generation protocols in the definition of $q_{{\sf link}}$, where it represents the probability of generating one successful ebit in one time slot.
Based on the model in~\cite{dahlberg2019link, quantum-queueing-delay}, the success probability $q_{{\sf link}}$ is defined as:
\begin{equation}
    \label{eq:q_e}
    q_{{\sf link}} = 1- \left[1 - q_{\sf gen} \left(1 - q_{\sf chan}\right)\right] ^ {N_{\sf attp}},
\end{equation}
where $q_{\sf gen}$ is the generation efficiency of entanglement source, and $q_{\sf chan}$ is the channel loss probability.

\begin{figure}[t]
\centering
\includegraphics[width=0.4\textwidth]{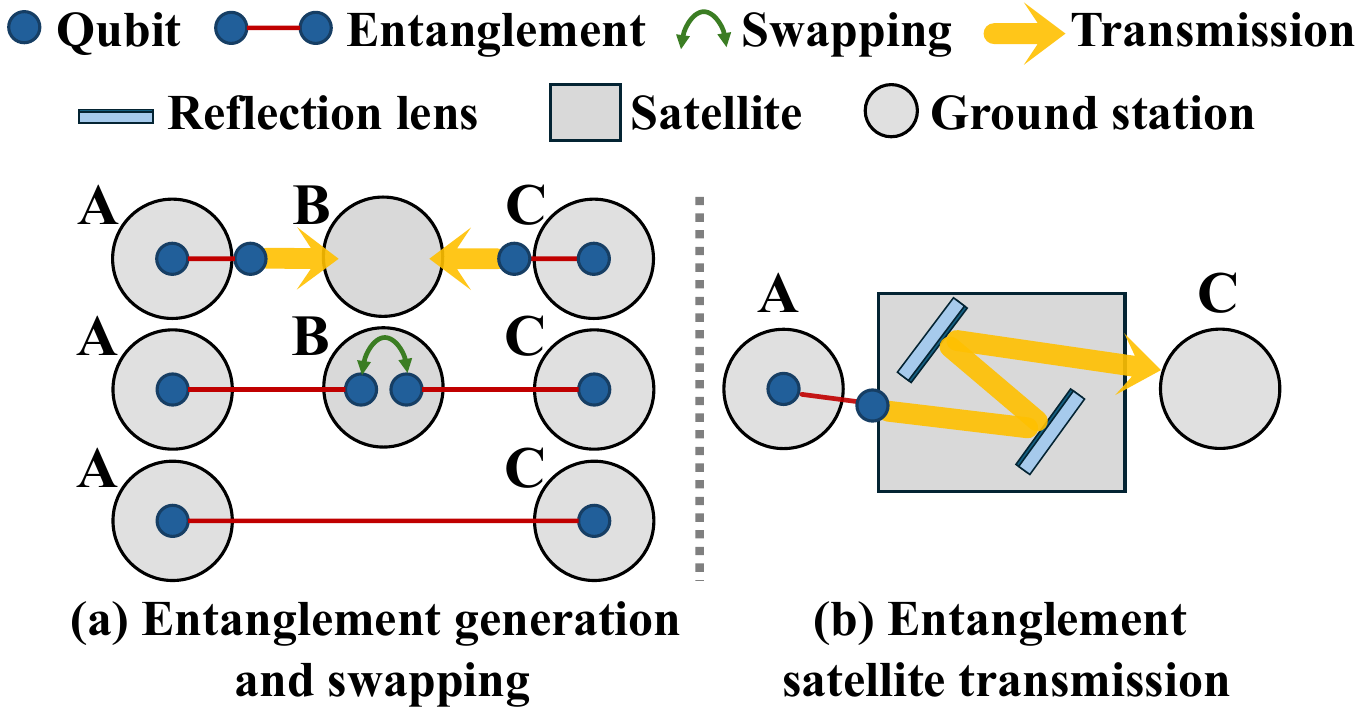}
\caption{\smallfont Entanglement generation, swapping versus direct transmission (using reflection lenses as an example).}
\label{fig:operations}
\end{figure}

\noindent\textbf{Entanglement swapping:}
This operation is performed when a quantum repeater receives one photon from each of two entangled pairs with different parties, and performs a local operation and classical communication (LOCC) procedure with one or two of the parties to establish a remote entanglement between them.
As in Fig.~\ref{fig:operations}(a), a Bell state measurement (BSM) is first performed between the two local photons at the repeater, which turns the two remote photons into one of the four Bell states.
One of the holders then performs a local one-qubit operation on its local photon based on the BSM result sent via a classical channel, correcting the two remote photons into the desired state.
This process creates an entanglement between two remote photons without interacting with each other.

In practice, entanglement swapping is limited by both theoretical and practical factors.
Theoretically, a BSM via linear optics yields the highest efficiency but suffers from a fundamental limit of up to $50\%$ success probability without auxiliary photons~\cite{calsamiglia2001maximum}, and using auxiliary photon pairs to boost the success probability to close to unity incurs an exponentially large number of auxiliary photons and photon detectors~\cite{bayerbach2023bell}.
In practice, any quantum operation (including swapping gates, detectors and/or auxiliary photons) incurs additional noise that further decreases the swapping success probability.
To model these effects, we regard entanglement swapping as a \emph{probabilistic process}, with a success probability $q_{\sf swp}$ that depends on the noise process and device setup on each repeater node.

\subsection{Entanglement Distribution via Ground/Satellite Networks}
\noindent
\noindent\textbf{Ground repeater network:}
A ground repeater network is built with geo-distributed repeater nodes that connect to users and each other via ground fiber links.
Entanglements are distributed between a pair of users via generation and swapping at a sequence of repeaters that form an entanglement path in the network.
Given that the ground network is static, the end-to-end entanglement distribution rate (EDR) depends on the links' entanglement generation rates and the success probabilities on nodes and links along the path, and is limited by exponential decay of the generation success probability over distance and the swapping success probability over number of hops. 
For ground fiber, the success probability $q_{\sf chan}$ is modeled with the fiber length $L$ and a amplitude damping factor $\gamma$ as~\cite{dahlberg2019link}:
\begin{equation}
    \label{eq:t_loss}
    q_{\sf chan} = 1 - 10^{-\frac{L\times \gamma}{10}}.
\end{equation}

\noindent\textbf{Satellite-assisted passive optical network:}
The satellite network utilizes direct line-of-sight between ground stations and satellites (called ground-satellite links or GSLs) and among satellites (called inter-satellite links or ISLs) to directly transmit entangled photons.
As shown in Fig.~\ref{fig:operations}(b), each satellite is equipped with adjustable lenses to reflect/refract and direct incoming photons to the next satellite or ground receiver.
A sequence of satellite-mounted lenses forms a direct lightpath from sender to receiver, and photons experience the most loss during uplink/downlink through the atmosphere or being reflected/refracted at lenses, and negligible absorption loss in-flight due to the near-vacuum environment~\cite{goswami2023satellite}.
The number of lightpaths that a satellite can form depends on the number of lenses and/or their steering angles, and the loss over each path depends mostly on the set of lenses of each satellite.

\noindent\textbf{Dynamic satellite topology:}
The satellite network is driven by the orbital dynamics of the satellites.
For LEO satellites, the high orbital velocity relative to the Earth's surface causes frequent disruptions of the GSLs~\cite{li2021internet,lai2023achieving}.
Meanwhile, the ISL topology remains nearly static in widely used constellations (such as the +Grid topology)~\cite{satellite_simulator}.
For quantum communication, this effect causes provisioned lightpaths to be frequently disrupted, such that new satellite lightpaths need to be established between ground stations.
On the other hand, because the orbital dynamics follow predictable patterns, future topology changes can be reliably predicted and taken into account when provisioning lightpaths.
Overall, orbital dynamics pose a major challenge that satellite-assisted quantum communication needs to address, and has been tackled in existing work~\cite{nitish_satellite_2022,chang2023entanglement_sat} for the single-satellite double-downlink distribution scheme.

\noindent\textbf{Hybrid ground-satellite entanglement distribution network:}
Given the pros and cons of both the ground repeater and the satellite-assisted passive optical networks, our proposal is a hybrid architecture that combines these two networks to achieve the best entanglement distribution efficiency at a global scale.
\begin{figure}[t]
\centering
\includegraphics[width=0.33\textwidth]{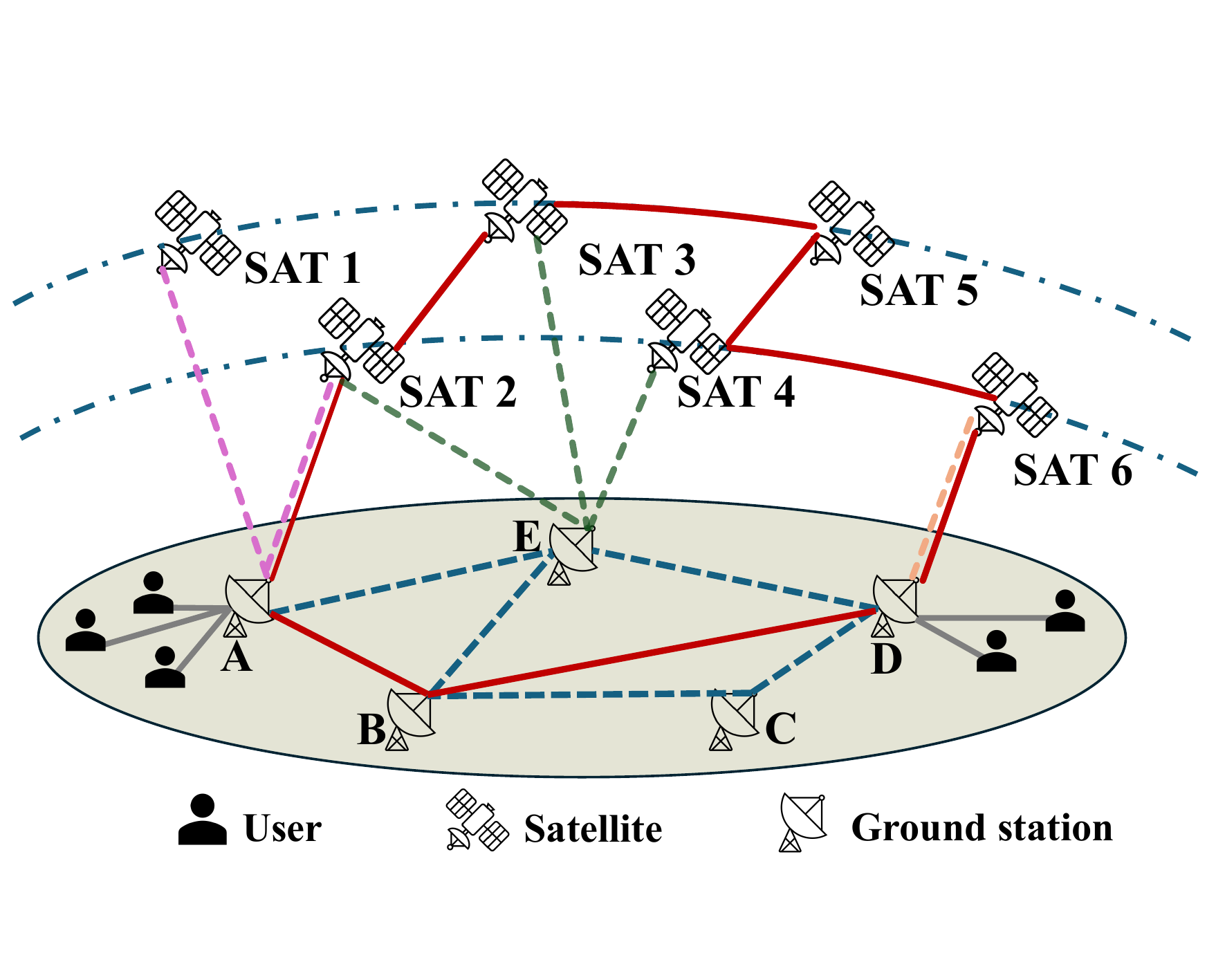}
\caption{\smallfont Entanglement distribution among $6$ ground stations and $6$ satellites with SD pair $AD$ connected to multiple users.}
\label{fig:satellite}
\end{figure}
As shown in Fig.~\ref{fig:satellite}, the dynamic satellite network is configured to establish long-distance direct lightpaths between major ground stations in the world, such as around major cities or critical infrastructure.
Complementary to the satellite network, ground fibers can interconnect ground stations, quantum repeaters as well as end users in proximity, such as via a fiber-based Quantum Metropolitan Area Network (QMAN)~\cite{chicago-quantum-MAN}.
Due to orbital dynamics, the satellite lightpaths need to be dynamically reconfigured to ensure optimal entanglement distribution across the globe, and entanglement distribution over the ground repeater network (augmented by satellite lightpaths) needs to take into account the dynamic supply of entanglement streams from space and the dynamic demand changes from ground users.
In the next sections, we formulate this problem and propose an algorithmic framework to solve it.

\section{Network Model and Problem Statement}
\label{section:ps}
\subsection{Quantum Network Model}
\noindent
Formally, we consider a quantum network $G = (V, E)$ as an undirected graph consisting of a ground repeater network $G^{\sf g} = (V^{\sf g}, E^{\sf g})$ with ground stations (equipped with quantum repeaters) interconnected by optical fibers and a satellite network $G^{\sf s} = (V^{\sf s}, E^{\sf s})$ with LEO satellites in multiple orbits.

Each ground station (called GS) $v^{\sf g} \in V^{\sf g}$ has a swapping success probability $q^{\sf g}_v \in (0,1]$. 
Each fiber link $e^{\sf g} \in E^{\sf g}$ has a capacity $c^{\sf g}_e \in \mathbb{Z}^+$, denoting the number of channels that can be attempted for ebit generation; $\mathbb{Z}^+$ denotes the positive integer set, and a generation success probability $q^{\sf g}_e \in (0,1]$.
Each satellite (called SAT) $v^{\sf s} \in V^{\sf s}$ has a capacity $c^{\sf s}_v$ to denote how many sets of lenses can be used to form lightpaths for direct photon transmission, and has a transmission success probability $q^{\sf s}_v \in (0,1]$ and the set $E^{\sf s}$ contains ISLs between satellites.

Additionally, we define GSLs as $E^{\sf gs}$ to represent the uplinks and downlinks between GSs and SATs.
Each GSL has a photon survival probability $q^{\sf gs}_e$ that is dominated by the absorption loss through the atmosphere; the uplink and downlink losses are typically asymmetric~\cite{goswami2023satellite}.
A satellite lightpath between two ground locations $g_1, g_2 \! \in \! V^{\sf g}$ is defined by a sequence of GSLs and ISLs, $p = (e_0, e_1, \dots, e_{\varkappa+1})$, where $e_0, e_{\varkappa+1} \in E^{\sf gs}$ and $e_i \! \in \! E^{\sf s}$ for $1 \leq i \leq \varkappa$; let the sequence of satellites along the lightpath be $v^{\sf s}_1, \dots, v^{\sf s}_\varkappa$.
The capacity of a satellite lightpath is decided by the joint optical bandwidth of all GSLs and ISLs plus that of the entanglement source, which we define as a uniform constant $\alpha$.
The end-to-end entanglement distribution success probability of a lightpath $p$ is the product of success probabilities of GSLs and satellites: $q(p) = q^{\sf gs}_{e_0} \cdot q^{\sf gs}_{e_{\varkappa+1}} \cdot \prod_{i=1}^{\varkappa} q^{\sf s}_{v_i}$.

The quantum network $G$ is constructed with the above two networks plus the GSLs: $V = V^{\sf g} \cup V^{\sf s}$ and $E = E^{\sf g} \cup E^{\sf s} \cup E^{\sf gs}$.
However, because of orbital dynamics, the set of GSLs can vary from time to time.
We thus define $E^{\sf gs}(t)$ as the set of GSLs available at time $t$, and $G(t)$ as the snapshot of the quantum network at time $t$.
The ISLs are assumed to be static through time following existing LEO constellation topologies~\cite{bhattacherjee2019network,lai2023achieving}.

Following existing work~\cite{zeng2022multi,panigrahy2022capacity}, we assume entanglement distribution is scheduled in unit time slot, and a central controller controls entanglement distribution between ground stations utilizing either fiber or satellite lightpaths.
Note that the entanglement distribution time slots are in a different time scale than the satellite dynamic topology changes.
Each time slot has two phases described below and depicted in Fig.~\ref{fig:operations}:
\begin{enumerate}
\item In Phase 1, entanglement generation is attempted along each fiber link $e$ or satellite lightpath $p$ between $m, n$, generating remote ebits with probability $q^{\sf g}_e$ or $q(p)$.
\item In Phase 2, for node pair $mn$ and intermediate node $k$ where ebits are available between both $mk$ and $kn$, node $k$ can attempt to perform swapping with success probability of $q^{\sf g}_k$ to establish ebits between $m$ and $n$ by using pairs of ebits between $mk$ and $kn$ in ground stations.
\end{enumerate}

We use $mn$ to interchangeably denote an \emph{unordered} node pair $\{ m, n \}$ for $m, n \!\in\!\! V^{\sf g} \cup V^{\sf s}$, and hence $mn\!=\!nm$.
Our model can be easily extended to the asynchronous model with repeaters, memories, and dynamic satellites.
Ebits can be temporarily stored in memories to wait for swapping with other ebits, and these phases could be executed in parallel because of the independent probabilistic operations in each phase.

\subsection{Commodities and Entanglement Demands}
\noindent

\noindent
We consider entanglement distribution between pairs of ground stations\footnote{Our model can be trivially extended to distributing between users, but the problem size will become very large in a global-scale quantum network.}. 
A pair of ground stations $sd$ is called a \emph{commodity}, and we use the traditional terminology of a source-destination (SD) pair even though the pair is unordered.
The set of all commodities in a graph is denoted by $\mathcal{U} \subseteq  V^{\sf g} \times V^{\sf g}$.

Commodities request remote entanglement distribution to support the quantum applications of their end users, and their requested demands can change over time.
Thus, for each commodity $i \in \mathcal{U}$, we define its entanglement demand as $z_i^t \ge 0$ for time $t$, denoting the requested \emph{expected EDR} between the two ground stations, we also denote demand set for commodity $i$ in all timeslots as $Z_i = \{ z_i^t \,|\, t \}$.
An intuitive way to represent such demands is by defining a sequence of \emph{demand matrices} between all pairs of ground stations over time.

The goal of the hybrid quantum network is to meet as many demands of commodities as possible by making real-time decisions to accommodate both topology and demand changes, which include provisioning satellite lightpaths and scheduling ground-satellite entanglement distribution.
\subsection{Problem Statement}
\noindent
Based on the network model, the quantum network operates on two timescales.
At a large timescale (consistent with the topology changing frequency), the network needs to provision inter-GS lightpaths for long-distance direct transmission of entangled photons.
At a small timescale (such as per second), the GSs utilize existing fiber or satellite lightpaths to generate elementary ebits and optionally perform swapping to extend or redistribute entanglements to meet demands.
We formally define the two-stage problem of lightpath provisioning and entanglement distribution in a satellite-assisted quantum network.

\noindent\textbf{Lightpath provisioning (LPP):}
The LPP problem aims to form a set of satellite lightpaths $P(t) = \{ p \}$ that interconnect ground stations over time.
Each $p$ sustains entangled photon transmission until either the uplink or downlink GSL is disrupted by orbital movement.
The lightpaths can be constructed over time when new GSLs become available.
Since lightpath provisioning takes coordination and commanding time due to satellites' movement, this step is done agnostic to the actual demands of commodities.
The goal is to maximize the total expected EDR for all commodities through these lightpaths.

\noindent\textbf{Entanglement distribution (EDT):}
Given pre-configured lightpaths and ground fibers, the EDT problem aims to establish entanglements across ground station pairs to meet the demands of users in real-time.
An EDT plan decides (1) the entanglement paths each consisting of fiber links, lightpaths and zero or more repeaters between an SD pair, and (2) the order of swapping across repeaters to yield the expected maximum EDR.
An EDT plan can be executed until either the demand changes, or when network topology changes affect a lightpath involved in the plan.
Note that the formulation of the EDT plan depends on the actual entanglement distribution protocol used, and we describe an optimal formulation that can be implemented with 
a probabilistic protocol in Sec.~\ref{sec:edt}.

Fig.~\ref{fig:satellite} is an example of the above processes.
The ground stations $A$ and $D$ have GSLs with one or more satellites, and one satellite such as $E$ has GSLs with multiple ground stations.
Between $AD$, there is an available satellite lightpath $2$-$3$-$5$-$4$-$6$ provisioned by solving the LPP problem.
On the ground, there is an additional entanglement path $A$-$B$-$D$ with swapping at ground station/repeater $B$.
Both paths can be utilized to solve the EDT problem and meet users' demands.
Entanglements distributed over a satellite lightpath can also be used to form entanglement paths for further swapping to serve end users only connected to ground stations $A$ and $D$.



\section{Two-stage Algorithm Design for Entanglement Distribution in QuESat}
\label{section:alg}
\noindent
\subsection{Satellite Lightpath Provisioning}
\noindent

\subsubsection{Problem Formulation} 
\noindent
We first define a mixed-integer linear programming (MILP) formulation for the LPP problem.
Let $T$ be the maximum planning period such as one orbital cycle or one day.
For any timeslot $t \in T$,
we define $\bar s_i^t, \bar d_i^t \in V^{\sf s}$ as the satellites that ground stations $s_i$ and $d_i$ are connected to at time $t$ by GSLs and define binary path selection variable $x_i^t(uv) \!\in\! \{ 0, 1\}$ to indicate if a lightpath for commodity $i$ utilizes an ISL $uv \!\in\! E^{\sf s}$ at time $t$.
The lightpath must originate from and end at satellites that the SD pair of ground stations have GSLs with.
Since a lightpath can sustain until one of the GSLs is disrupted, and should indeed do so to avoid excessive lightpath reconfiguration overheads, we define variable $x_i^{m^t_i}(uv)$ to replace $x_i^t(uv)$, where $m^t_i$ is the number of topology changes that have affected commodity $i$'s GSLs by time $t$ and instead use $\bar s_i^{m^t_i}, \bar d_i^{m^t_i}$ to represent pair of satellites that ground stations $s_i$ and $d_i$ are connected to if topology changes implied by $m^t_i$.
This indicates that the lightpath for a commodity does not change until its GSLs change.
Each lightpath occupies one set of lenses on each satellite, subject to the maximum number of sets carried by the satellite.

The goal of LPP is to maximize the total throughput of lightpaths provisioned between all SD pairs.
We further define variables $f_i^{m^t_i}(u,v) \in \mathbb{R}^*$ to denote the expected entanglement flow for commodity $i$ carried on ISL $uv$.
The expected flow incurs loss when passing through each satellite lens, and is bounded by the overall lightpath capacity $\alpha$ from the entanglement source.
We formulate the LPP problem as follows:
\begin{subequations}
\label{fml:ilp_2}
\begin{equation}
    \text{maximize} \quad \sum_{t\in T}\nolimits \sum_{i\in \mathcal{U}} \nolimits\eta_i^{m^t_i}   \tag{\ref{fml:ilp_2}}
\end{equation}
\text{subject to} 
\begin{equation}
    \begin{aligned}
    \label{fml:ilp-2-cons2}
     &
     f_i^{m^t_i}(u,v) + f_i^{m^t_i}(v,u)  \le \alpha \cdot x_i^{m^t_i}(uv), \\
     &
    \quad\quad \quad\quad \quad\quad  \quad\quad 
     \forall uv \in E^{\sf s}, \forall i  \in \mathcal{U},  \forall t \in T;
     \end{aligned}
\end{equation}
\begin{equation}
    \begin{aligned}
    \label{fml:ilp-2-cons1}
    &
    \sum_{u:uv \in E^{\sf s}}\sum_{i\in \mathcal{U}}\ \!\!x_i^{m^t_i}(uv) \le c_v^{\sf s}, \quad\forall v \in V^{\sf s}, \forall t \in T;
    \end{aligned}
\end{equation}
\begin{equation}
    \begin{aligned}
    \label{fml:ilp-2-cons3}
    &
    q^{\sf s}_v\sum_{\mathclap{v: uv\in E^{\sf s}}} f_i^{m^t_i}(u,v) - \sum_{\mathclap{v: uv\in E^{\sf s}}}f_i^{m^t_i}(v,u) = 0,\\
    &
    \quad\quad \quad\quad \quad\quad 
    \forall i  \in \mathcal{U}, \forall v \in V^{\sf s} \setminus \left\{ \bar s_i^{m^t_i}, \bar d_i^{m^t_i} \right\},  \forall t \in T;
     \end{aligned}
\end{equation}
\begin{equation}
    \begin{aligned}
    \label{fml:ilp-2-cons5}
    &
    q^{\sf s}_v\sum_{\mathclap{v: uv\in E^{\sf s}}}f_i^{m^t_i}(u,v) - \sum_{\mathclap{v: uv \in E^{\sf s}}} f_i^{m^t_i}(v,u) =\eta_i^{m^t_i}, \\
    &
    \quad\quad \quad\quad \quad\quad  \quad\quad 
    \forall i  \in \mathcal{U}, v = \bar d_i^{m^t_i}, \forall t \in T;
 \end{aligned}
\end{equation}
\begin{equation}
    \begin{aligned}
    \label{fml:ilp-2-cons6}
    &
     f_i^{m^t_i}(u,v), \eta_i^{m^t_i} \ge 0, x_i^{m^t_i}(uv) \in \{ 0, 1\}, \\
     &
      \quad\quad  \quad\quad  \quad\quad  \quad\quad 
     \forall uv \in E^{\sf s},\forall i  \in \mathcal{U}, \forall t \in T.
     \end{aligned}
\end{equation}
\end{subequations}
\noindent\textbf{Explanation:} Program~\eqref{fml:ilp_2} aims to maximize the total throughput for all lightpaths in all timeslots while considering the limited pairs of lenses in each satellite and the multiplicative loss when photons pass through each set of lenses.
Program~\eqref{fml:ilp_2} only updates the solution when GSLs of any commodity change, which is implied by the index function $m^t_i$.
Constraint~\eqref{fml:ilp-2-cons2} couples the entanglement flow variables $\{f_i^{m^t_i}(u,v)\}$ and the path selection variables $\{x_i^{m^t_i}(uv)\}$, enforcing that photon traffic is only carried on provisioned lightpaths and bounded by the source/link capacity $\alpha$.
Constraint~\eqref{fml:ilp-2-cons1} limits the number of lightpaths through a satellite by its equipped sets of lenses. An ISL used in a lightpath occupies a set of lenses on both satellites, and thus counts towards both capacities $c^{\sf s}_u$ and $c^{\sf s}_v$.
Constraints~\eqref{fml:ilp-2-cons3}--\eqref{fml:ilp-2-cons5} jointly serve two purposes: (1) \emph{flow conservation:} they ensure that the $\{f_i^{m^t_i}(u,v)\}$ variables, and hence also the $\{x_i^{m^t_i}(uv)\}$ variables as well by Constraint~\eqref{fml:ilp-2-cons2}, form lightpaths for SD pairs; (2) \emph{loss accumulation:} they enforce a loss corresponding to the survivable probability $q^{\sf s}_v$ when the photon traffic passes through each set of lenses on satellite $v$, thus discounting the final received flow at the receiver by the product of survival probabilities of all satellite lenses along each lightpath.
The throughput $\eta_i^{m^t_i}$ achieved by each SD pair's lightpath is the flow received at the receiver side, discounted by the cumulative loss.
As a side product, these constraints also cancel any loop that may form since each satellite's lenses strictly decrease the final received flow.

\subsubsection{Randomized Rounding-based Algorithm}
\noindent
\begin{algorithm}[t]
\caption{\mbox{LPP with Randomized Rounding (LPP-RR)}}
\label{a:randomized}
\KwIn{Satellite topology $G^{\sf s} = (V^{\sf s},E^{\sf s})$, commodity set $\mathcal{U}$, topology change indices $\{ m_i^t \}$}
\KwOut{Lightpaths $P$ selected for provisioning}
$P \leftarrow \emptyset$\;
Relax each binary integer $x_i^{m_i^t}{(uv)}$ to $\hat{x}_i^{m_i^t}{(uv)} \in [0, 1]$\;
Solve relaxed Program~\eqref{fml:ilp_2} to obtain optimal solution $\{\hat x_i^{m^t_i}(uv)\}$ for all commodities $i \in \mathcal{U}$\;
Decompose $\{ \hat x_i^{m^t_i}(uv) \}$ to a set of candidate lightpaths $P^{\sf cand}$ each with fractional path selection value $\hat{x}^{m^t_i}_p$\;\label{alg-r:2}
\For{all lightpaths $p \in P^{\sf cand}$}{
    With probability $\hat{x}_p^{m^t_i}$, 
    let $P \leftarrow P \cup \{ p \}$;
    }\label{alg-r:randomized}
\Return{all selected lightpaths $P$.}
\end{algorithm}
\noindent
Program~\eqref{fml:ilp_2} is a MILP that can be reduced from the edge-disjoint path, which is NP-hard~\cite{mip}.
To tackle this, we first develop a randomized rounding-based algorithm to approximate the optimal solution, as shown in Algorithm~\ref{a:randomized}.
The algorithm relaxes the integral constraint in $\{ x_i^{m^t_i}(uv) \}$ and lets it take continuous values in $[0, 1]$, and then solves the relaxed linear program (LP) to obtain optimal fractional solutions $\{\hat x_i^{m^t_i}(uv)\}$.
After that, it runs a flow-decomposition algorithm to decompose the $\{ \hat x_i^{m^t_i}(uv) \}$ into a set of candidate lightpaths $P^{\sf cand} = \{p \,|\, p \in P_i^{m_i^t}, i \in \mathcal{U}\}$ where $P_i^{m_i^t}$ is the paths set for commodity $i$ at timeslot $t$ with the corresponding index function $m_i^t$,  each with a fractional path selection value $\hat{x}^{m^t_i}_p$.
Finally, the algorithm randomly selects each candidate lightpath with probability $\hat{x}_p^{m^t_i}$, and returns the set of selected lightpaths as the solution.

The randomized rounding algorithm enjoys a probabilistic guarantee in its optimality and constraint satisfaction.
We assume $ALG$ is the solution obtained by Algorithm~\ref{a:randomized}, and $LP$ is the solution obtained by the relaxed version of Program~\eqref{fml:ilp_2}. Let the throughput output by Program~\eqref{fml:ilp_2} be $SOL_{LP}$ and the throughput output by Algorithm~\ref{a:randomized} be $SOL_{ALG}$.
The following theorems follow directly from applying the Chernoff bound and the Hoeffding's inequality to the objective value (sum of lightpath's expected EDRs) and each node capacity constraint in Eq.~\eqref{fml:ilp-2-cons1}, respectively, and we omit their proofs while referring the readers to similar proofs in~\cite{liu2019constant,dubhashi1998concentration,kodialam2003dynamic}.
\begin{theorem}
\label{th:1}
Let $\sigma = q_{\sf max} / q_{\sf min}$ where $q_{\sf max}, q_{\sf min}$ are the maximum and minimum end-to-end survivable probability of any lightpath in $G^{\sf s}$. Then for any $\epsilon \in (0, 1)$, we have:
$\Pr[SOL_{ALG} <(1-\epsilon) \cdot  SOL_{OPT}]  \le e^{-\epsilon^2 \cdot \sigma \cdot SOL_{OPT} / 2}$.
    \myendbox
\end{theorem}
\begin{theorem}
\label{th:2}
    For a satellite $v$ with lense capacity $c^{\sf s}_v$, at any time $t$, let $\hat x_{i, v} = \sum_{u} \hat x_i^{m^t_i}(uv)$ be the total fractional path selection value for SD pair $i$ on node $v$, and let $\hat x_{v} = \sum_{i} \hat x_{i, v}$.
Define $\varsigma = \max_i \{ \hat x_{i,v} / (\hat x_{v} \cdot c^{\sf s}_v) \}$.
The probability that Algorithm~\ref{a:randomized} exceeds the node capacity $c^s_v$ at time $t$ by a factor of $1 + \varsigma \sqrt{ 2 \log |V^{\sf s}| \cdot |\mathcal{U}|}$ is upper bounded by $|V^{\sf s}|^{-2}$.
    \myendbox
\end{theorem}
Based on Theorem~\ref{th:2}, randomized rounding may cause capacity violation when the number of lightpaths exceeds the number of lenses on a satellite.
This can be resolved by applying a pruning procedure to remove excess lightpaths, as shown later.
Since Program~\eqref{fml:ilp_2} is a packing problem, the randomized rounding algorithm can further be derandomized with the method of conditional probabilities invented by Raghavan~\cite{raghavan1988probabilistic}.
To keep our solution simple, we instead describe a deterministic rounding algorithm that fully respects the node capacity constraints and shows good practical performance.

\subsubsection{Deterministic Rounding-based Algorithm}
\noindent
We design a simple deterministic rounding algorithm in Algorithm~\ref{a:heuristic} inspired by~\cite{alicherry2012network,kar2003routing}.
The algorithm follows the same initial steps as Algorithm~\ref{a:randomized} by first solving the relaxed linear program and obtaining fractional solutions $\{\hat x_i^{m^t_i}(uv)\}$, and then obtaining a sequence of candidate lightpaths $P^{\sf cand} = \{p \,|\, p \in P_i^{m_i^t}, i \in \mathcal{U}\}$ each with fractional path selection value $\hat{x}^{m^t_i}_p$.
Then, the algorithm selects all lightpaths whose fractional path selection values are over a certain threshold $\Delta$.
To ensure feasibility, the algorithm further iterates over all satellite nodes, and for any node whose capacity constraint is exceeded, the algorithm removes an arbitrary path through it, until all capacity constraints are satisfied.
The algorithm returns a feasible solution to Program~\eqref{fml:ilp_2}.
We note that the last step of recovering to feasibility also applies to the output of the randomized rounding in Algorithm~\ref{a:heuristic}, and will employ this step for both algorithms to ensure feasibility in evaluation.
\begin{algorithm}[t]
\caption{\mbox{LPP with Deterministic Rounding (LPP-DR)}}
\label{a:heuristic}
\KwIn{Satellite topology $G^{\sf s} = (V^{\sf s},E^{\sf s})$, commodities $\mathcal{U}$, topology change indices $\{ m_i^t \}$, threshold $\Delta$}
\KwOut{Lightpaths $P$ selected for provisioning}
$P \leftarrow \emptyset$\;
Relax each binary integer $x_i^{m_i^t}{(uv)}$ to $\hat{x}_i^{m_i^t}{(uv)} \in [0, 1]$\;
Solve relaxed Program~\eqref{fml:ilp_2} to obtain optimal solution $\{\hat x_i^{m^t_i}(uv)\}$ for all commodities $i \in \mathcal{U}$\;
Decompose $\{ \hat x_i^{m^t_i}(uv) \}$ to a set of candidate lightpaths $P^{\sf cand}$ each with fractional path selection value $\hat{x}^{m^t_i}_p$\;
Let $P \leftarrow \{ p \in P^{\sf cand}
 \,|\, \hat x_p^{m^i_i} \ge \Delta \}$\;
\For{each satellite $v \in V^{\sf s}$}{
    \textbf{while} $\sum_{p \in P: v \in p} > c^{\sf s}_v$ \textbf{do} $P \leftarrow P \setminus \{ p \}$
    }
\Return{all selected lightpaths $P$.}
\end{algorithm}

\subsection{The Entanglement Distribution Problem}
\label{sec:edt}
\noindent
Given the satellite lightpaths provisioned based on the LPP algorithms and the ground fiber links, the EDT problem seeks to schedule entanglement distribution among commodities to meet their time-varying demands.
Since LPP is agnostic to actual demands, it may frequently happen that the provisioned lightpaths are insufficient to meet the demands of certain commodities, and so entanglement (re-)distribution is needed via additional ground entanglement generation and entanglement swapping by the repeater devices in ground stations.

Let $P_{mn}(t)$ be the set of lightpaths between ground stations $mn$ at time $t$, each with capacity $\alpha$ and success/survival probability $q(p)$.
We first construct an \emph{augmented ground repeater network} $G^{\sf g+}(t) = (V^{\sf g}, E^{\sf g+}(t))$ where $E^{\sf g+}(t) = \left(\bigcup_{mn} P_{mn}(t)\right) \cup E^{\sf g}$ contains both ground fiber links, and satellite lightpaths as additional virtual links between ground station pairs; each $G^{\sf g+}(t)$ is thus a multi-graph with possibly more than one link between every pair of nodes.

To solve the EDT problem, we extend an existing optimal single-commodity entanglement distribution formulation~\cite{Dai2020a} to multi-commodity entanglement distribution in a multi-graph.
Note that between different topology or demand change time slots, the EDT decisions are independent and centrally controlled by a controller.
This allows us to consider each time slot independently and drops the time index $t$ in the formulation below.
Define variable $g_{mn} \in [0, 1]$ as the elementary ebit generation ratio over the combined capacity of both ground link and all lightpaths between a ground station pair $m,n \in V^{\sf g}$.
Define variable $y^{mn}_{mk}$ as the swapping rate from $mn$-ebits to $mk$-ebits at repeater $n$ for $m, n, k \in V^{\sf g}$.
The optimal EDT problem for maximizing total EDR is formulated as follows:
\begin{subequations}
\label{fml:0}
\begin{align}
    \text{maximize} \quad \sum_{i \in \mathcal{U}}\nolimits \zeta_i \tag{\ref{fml:0}}  
\end{align}
\text{subject to} 
\begin{align}
    & 
    y^{mk}_{mn} = y^{kn}_{mn}, 
    \quad \forall k, m, n \in V^{\sf g};
    \label{mored:eq:1}
\end{align}
\begin{equation}
    \begin{aligned}
        I_{mn} - \Omega_{mn} = 
        \left\{
        \begin{aligned}
            & 0, && \text{if } mn \notin \mathcal{U}, \\
            & \zeta_i, && \text{if } mn = s_i d_i \in \mathcal{U}.\\
        \end{aligned}
        \right.
    \end{aligned}
    \label{mored:eq:2}
\end{equation}
\text{Here $I(mn)$ and $\Omega(mn)$ are defined as:}
\begin{equation}
\begin{aligned}
    & \quad
    I(mn) \triangleq \left(  q^{\sf g}_{mn} \cdot c^{\sf g}_{mn} + \sum_{p \in P_{mn}}\nolimits q(p) \cdot \alpha   \right) g_{mn}\\
    & \quad \quad \quad \quad \quad
    + \frac{1}{2} \cdot \sum_{{k \in V^{\sf g} \setminus \{m, n\}}}\nolimits(y^{mk}_{mn} + y^{kn}_{mn}) \cdot q^{\sf g}_k; \label{mored:eq:4}
\end{aligned}
\end{equation}
\begin{equation}
\begin{aligned}
    &  
    \Omega(mn) \triangleq \sum_{{k \in V^{\sf g} \setminus \{m, n\}}}\nolimits  \left( y^{mn}_{mk} + y^{mn}_{kn} \right);
\end{aligned}
\end{equation}
\begin{equation}
    \begin{aligned}
    &
        g_{mn} \in [0, 1], \, y^{mn}_{mk}, \, \zeta_i \ge 0, \forall k,m,n \in V^{\sf g},\forall i \in \mathcal{U}.\label{mored:eq:var}
    \end{aligned}
\end{equation}
%
%
\noindent\textbf{Explanation:} 
In the above, $I(mn)$ denotes the \emph{input} (established) ebits between $mn$ through entanglement generation (via ground link or satellite lightpaths) and/or swapping, and $\Omega(mn)$ denotes the \emph{output} (consumed) $mn$-ebits for swapping to generate ebits between other node pairs.
The objective is to maximize the overall EDR of all commodities. 
Constraint~\eqref{mored:eq:1} makes sure that every swapping between two node pairs $mk$ and $kn$ at node $k$ consumes the same number of ebits; 
Constraint~\eqref{mored:eq:2} requires all acquired ebits among repeaters to be used for further swapping, and for SD pair $i$, the acquired ebits should be no less than the distributed ebits.
The following theorem follows from Theorem~1 in~\cite{Dai2020a} showing optimality of the EDT formulation \emph{w.r.t.} the total EDR of all commodities:
\begin{theorem}
\label{th:opt}
Given quantum network $G^{\sf g+}$ with both fiber links and satellite lightpaths, the maximum achievable total EDR between all commodities by any entanglement distribution protocol is upper bounded by the optimal value of Program~\eqref{fml:0}.
\end{theorem}

Program~\eqref{fml:0} can also be modified to meet the demands of commodities as much as possible by adding constraint:
\begin{equation}
\label{extra:cons}
\zeta_i \le z_i, \quad \forall i \in \mathcal{U}.
\end{equation}
\end{subequations}

Other goals can also be formulated, such as strictly satisfying all demands while minimizing the maximum node/link utilization, minimizing distribution cost, or achieving fairness of commodities by changing the objective in Program~\eqref{fml:0} to:
\begin{equation}
    \max \min \{ \zeta_i / z_i \,|\, \forall i \in \mathcal{U} \}.
\label{extra:obj}
\end{equation}

The above formulations result in a linear program (or a convex program in the case of fairness) that a solver can efficiently solve.
Further, the solution can be fully implemented as a probabilistic protocol to achieve the same expected EDR in the long run.
Due to space limits, we omit the probabilistic protocol, which is a trivial extension in~\cite{fendi}.
Other entanglement distribution protocols can also be used in place of Eq.~\eqref{fml:0} proposed in existing work~\cite{zeng2022multi,zhao2024Routing}.
Since Eq.~\eqref{fml:0} is adopted to optimize for different objectives, we choose to use it instead of other algorithms for evaluating the advantage of QuESat.

\subsection{The QuESat Entanglement Distribution Framework}
\noindent
Based on the above building blocks, the QuESat entanglement distribution framework is shown in Algorithm~\ref{a:0}.
Periodically, such as per orbital cycle of the satellites, one of the LPP algorithms is called to command the dynamic provisioning of satellite lightpaths based on predictable GSL availability during the next cycle.
Afterwards, the network controller monitors changes in user demand patterns across ground stations/repeaters, and calculates the EDT solution prior to or in response to a predicted change.
The solution is then distributed to all ground stations/repeaters, which executes the probabilistic entanglement generation and swapping protocol distributedly based on the solution.
With both fiber links and satellite lightpaths, the architecture should provide high entanglement distribution rate both at short distances (by ground fiber) and long distances (by satellite), thereby achieving high-efficiency entanglement distribution across the globe.

\begin{algorithm}[t]
\caption{\mbox{QuESat Entanglement Distribution Framework}}
\label{a:0}
\For{each lightpath planning period $T$}{
    Call LPP-RR or LPP-DR to decide lightpaths $P$ for the next period\;
    \For{each topology change time slot $t \in T$}{
    Provision lightpaths according to $P$\;
    Generate augmented ground repeater network $G^{\sf g+}(t)$ based on $P$\;
    Compute the optimal EDT by solving Program~\eqref{fml:0} on $G^{\sf g+}(t)$\;
    }
    \For{each entanglement distribution time slot}{
    Generate elementary ebits via ground links and satellite lightpaths\;
    Perform swapping based on EDT plan.
    }
}
\end{algorithm}



\section{Performance Evaluation}
\label{sec:results}
\noindent
\subsection{Evaluation Methodology}
\begin{figure}[t]
\vspace{-1em}
\centering
\subfloat[Throughput comparison]
{\includegraphics[width=0.24\textwidth]{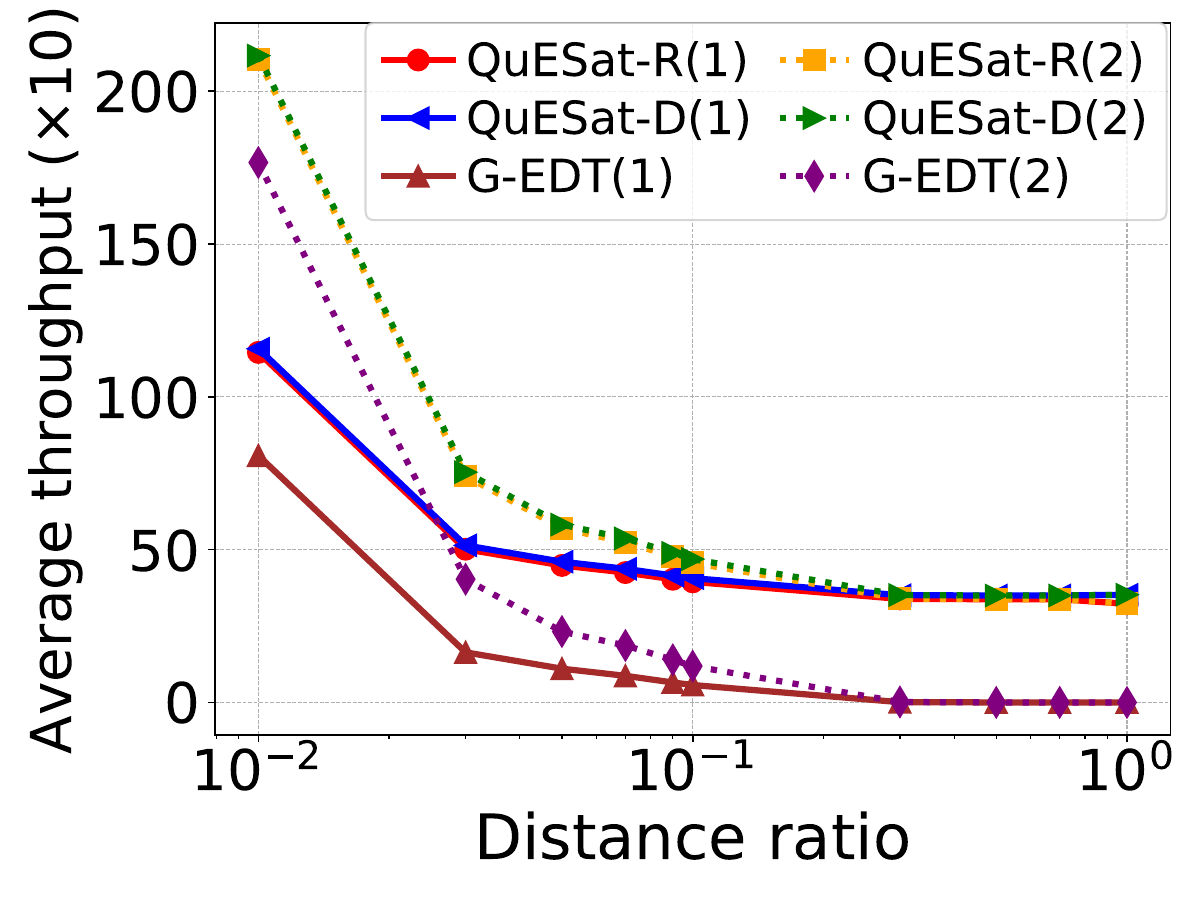}
\label{fig:distance:averagethroughput}}
\subfloat[Satisfaction ratio comparison]
{\includegraphics[width=0.24\textwidth]{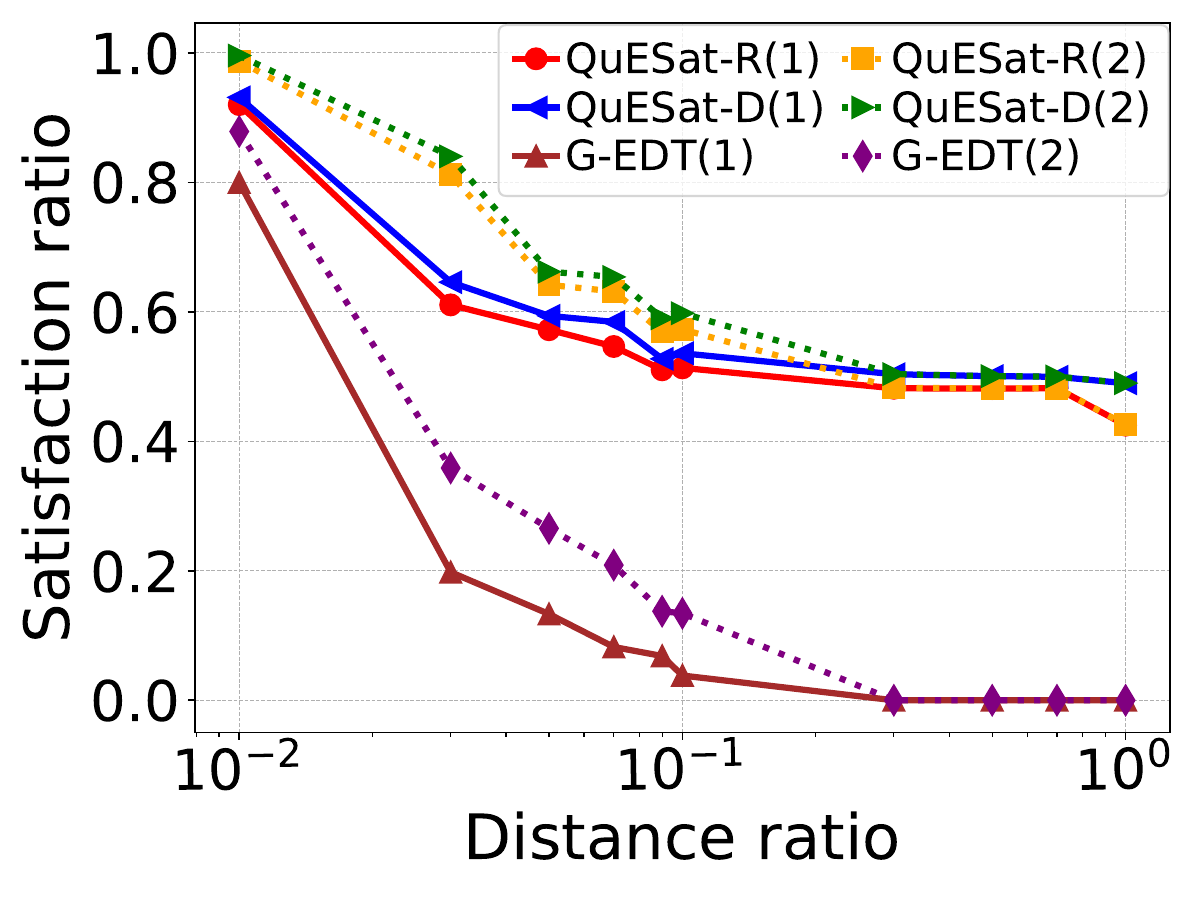}
\label{fig:distance:satisfication_ratio}}
\vspace{-2mm}
\caption{Impact of distance ratio for QuESat and G-EDT}
\label{fig:distance}
\end{figure}

\noindent
We simulated an LEO satellite constellation with $10$ ground stations located in major cities around the world.
The distances between ground stations were calculated by longitude and latitude provided in~\cite{satellite_simulator}.
Considering the advances in quantum memory with over $10$ seconds of coherence time, we set each entanglement distribution time slot to be $10$ seconds~\cite{nv_1_46s,nv_10}.
Following existing work~\cite{zhao2024Routing,shi2020concurrent}, each node had a swapping success probability uniformly sampled from $[0.85,0.98]$.
We assumed fiber links existed between every pair of ground stations.
One observation that we had was that computing fiber loss over the original distances between worldwide major cities resulted in almost zero success in entanglement generation, so we scaled the distance between ground stations by a factor of $0.1$ to simulate a much smaller region where fiber-based entanglement distribution can be more practical~\cite{fendi,Dai2020a}.
We explored the impact of distance by varying this factor in Fig.~\ref{fig:distance}.
The edge capacity was uniformly set as $10$ ebits-per-slot.

For satellite orbital dynamics, we used the SILLEO-SCNS simulator~\cite{satellite_simulator} to simulate an LEO constellation with $10$ planes and $15$ satellites per plane, and plane inclination was $96.9$ degree.
The constellation provided seamless coverage of all ground stations. 
The widely used +Grid topology was applied to build the ISLs, where each satellite had ISLs with the two neighbors in the same orbit and one nearest satellite in either neighboring orbit; this resulted in $4$ ISLs per satellite~\cite{starlink}.
We run each simulation for 24 hours, corresponding to about 15 orbital cycles per satellite.
We monitored GSL changes every $10$ minutes.
The ground-based entanglement source had a capacity of $10$ ebits-per-slot, same as the optical fibers. 
The uplink/downlink combined absorption loss was set to $90\%$ (with a survivable probability of $10\%$), while the loss per lens was uniformly sampled from $2\%$ to $5\%$ following~\cite{goswami2023satellite}.

We selected $15$ random SD pairs from the set of ground stations and generated demands following the gravity model~\cite{tune2015spatiotemporal}.
The demands changed every hour, based on randomly generated user population in $[70, 300]$ at each ground station with a total demand of $40000$.
Linear programs were solved by the Gurobi solver~\cite{gurobi}.
Results were averaged over 3 runs in the same setting.
The following baselines were compared:
\begin{itemize}
    \item \textbf{QuESat-D:} QuESat with LPP-DR and EDT with $\Delta\!=\!0.5$;
    \item \textbf{QuESat-R:} QuESat with LPP-RR and EDT;
    \item \textbf{G-EDT:} Entanglement distribution over only the ground fiber network with optimal EDT maximizing total EDR.
\end{itemize}
We simulated two scenarios: (1) The SD pairs only distributed and swapped entanglements among themselves without relying on additional ground stations as repeaters. (2) All $35$ ground stations were utilized as repeaters to serve demands of users.
We use ``(1)'' and ``(2)'' following an algorithm's name to denote the scenario in which the algorithm was evaluated.

The following metrics were used for evaluation.
The \textbf{\emph{average throughput}} measures the average EDR over all SD pairs.
The \textbf{\emph{satisfaction ratio}} measures the average ratio of the number of SD pairs whose demand can be satisfied among all SD pairs.

\subsection{Evaluation Results}
\begin{figure}[t]
\vspace{-1em}
\centering
\subfloat[Average throughput in Scenario(1)]
{\includegraphics[width=0.24\textwidth]{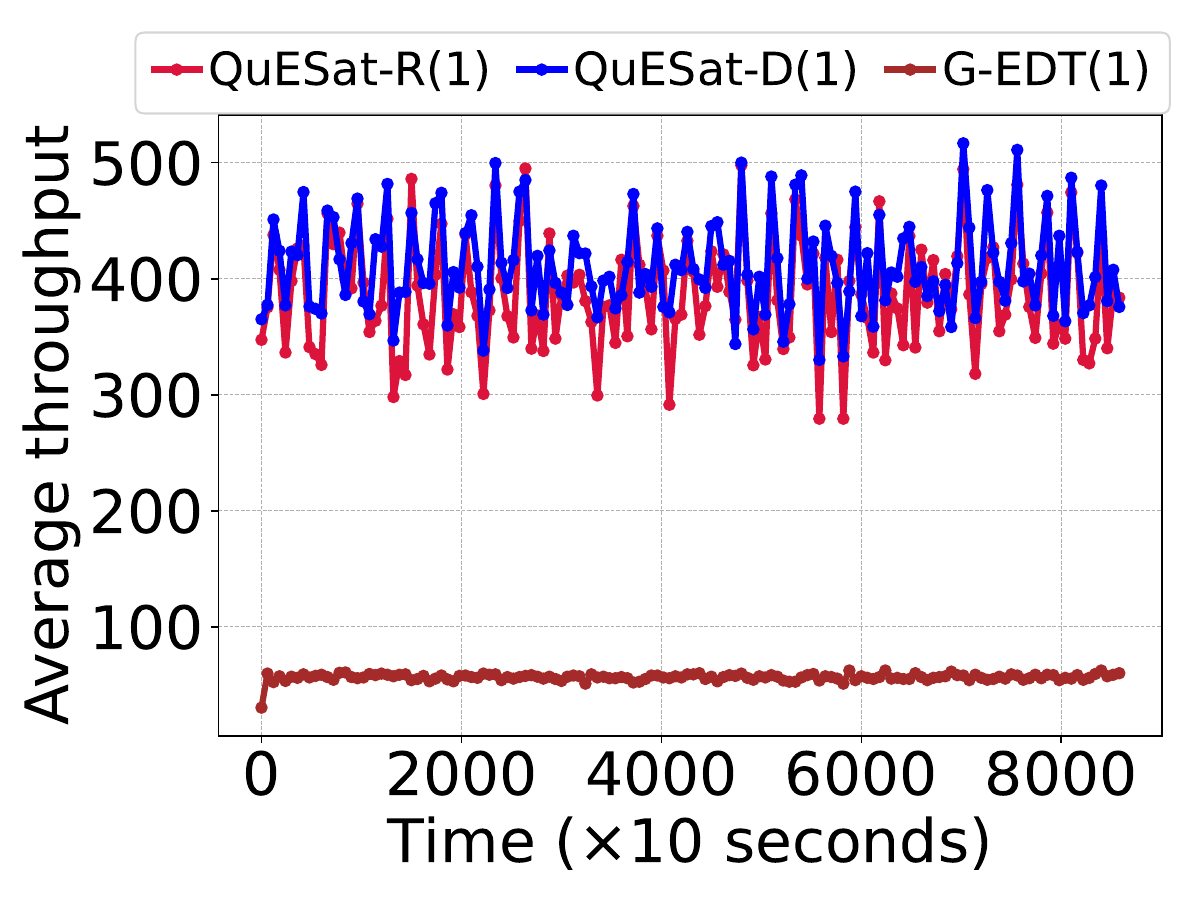}
\label{fig:time:average_throughput_commodities}}
\subfloat[Average throughput in Scenario(2)]
{\includegraphics[width=0.24\textwidth]{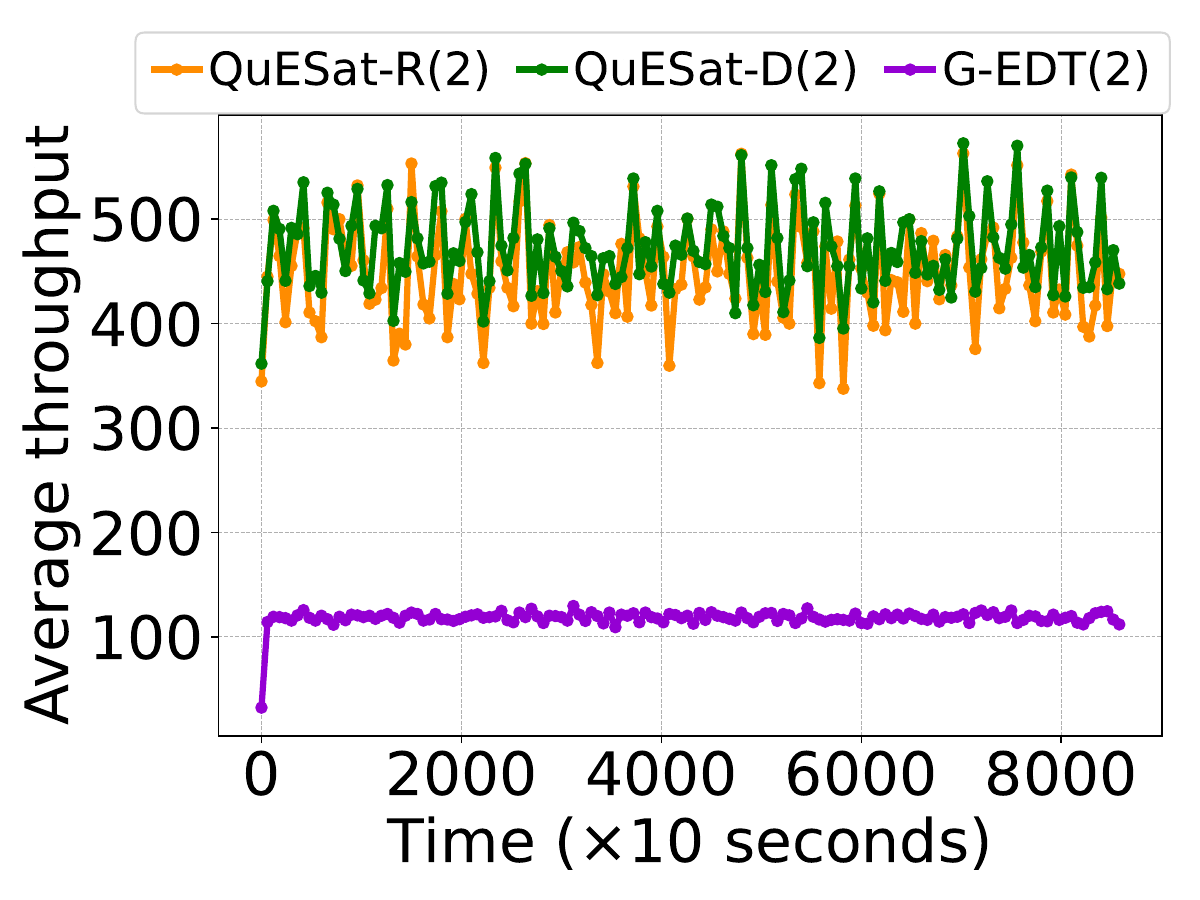}
\label{fig:time:average_throughput_fullgraph}}
\vspace{-2mm}
\caption{Comparison for average throughput over time}
\label{fig:time_throughput}
\end{figure}
\begin{figure}[t]
\vspace{-1em}
\centering
\subfloat[Satisfaction ratio in Scenario(1)]
{\includegraphics[width=0.24\textwidth]{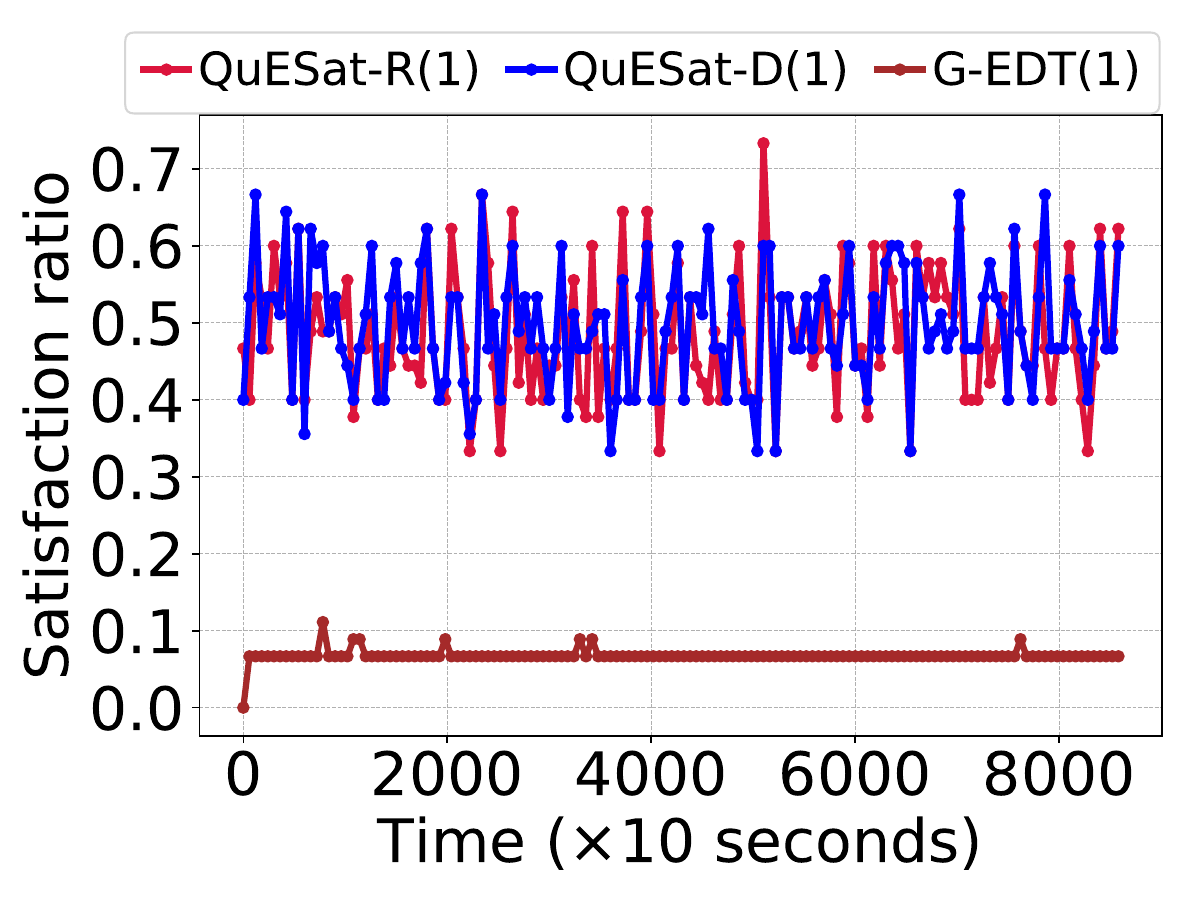}
\label{fig:time:satisfication_ratio_commodities}}
\subfloat[Satisfaction ratio in Scenario(2)]
{\includegraphics[width=0.24\textwidth]{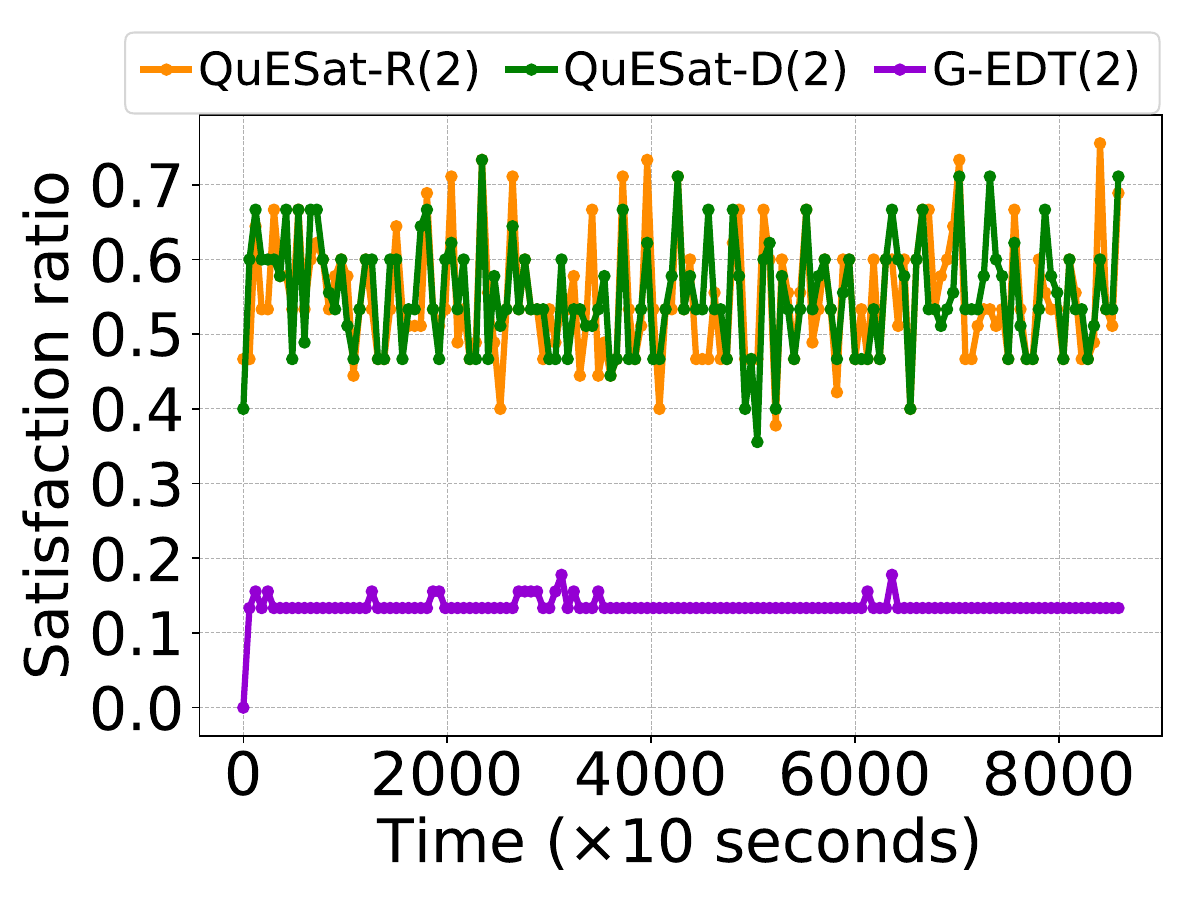}
\label{fig:time:satisfication_ratio_fullgraph}}
\vspace{-2mm}
\caption{Comparison for satisfaction ratio over time}
\label{fig:time_satis_ratio}
\end{figure}
\begin{figure}[t]
\vspace{-1em}
\centering
\subfloat[Throughput comparison]{\includegraphics[width=0.24\textwidth]{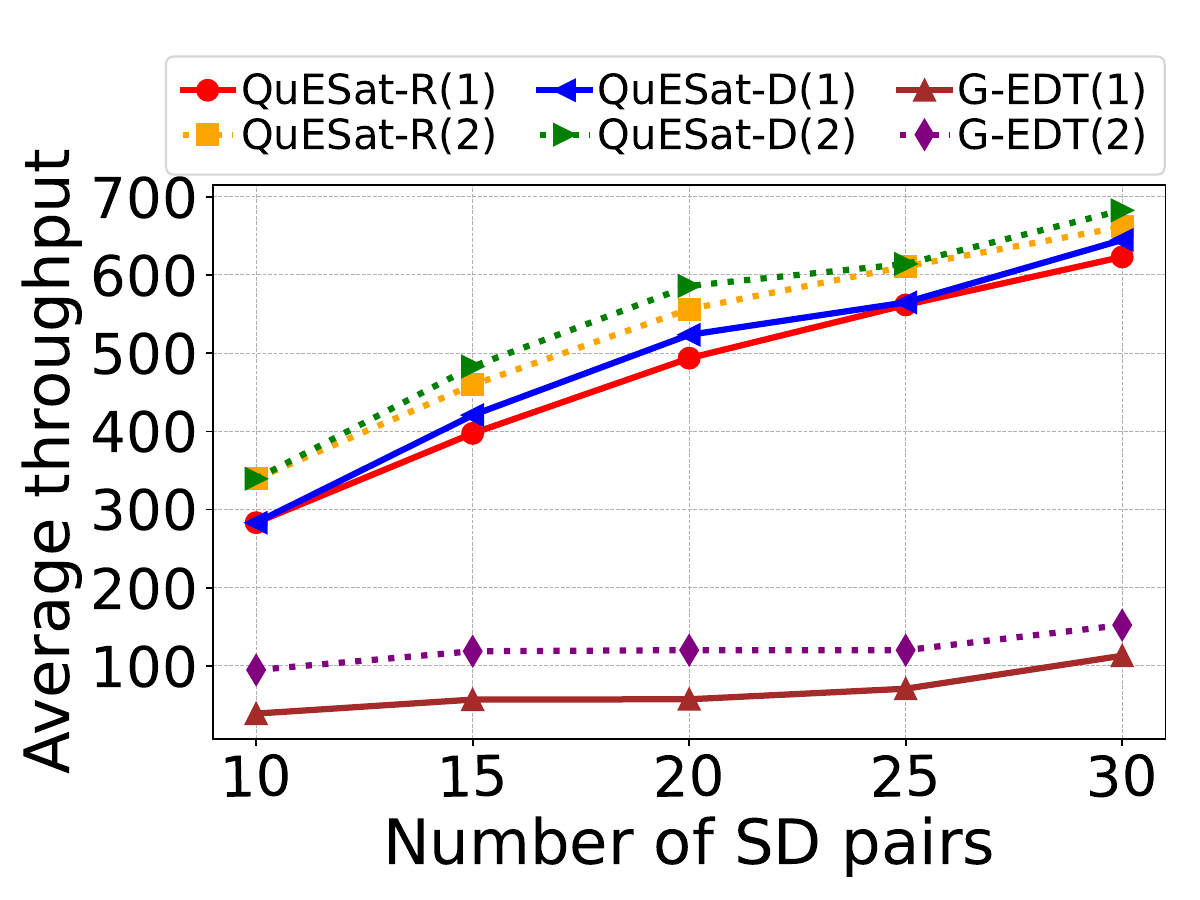}
\label{fig:pairs:averagethroughput}}
\subfloat[Satisfaction ratio comparison]
{\includegraphics[width=0.24\textwidth]{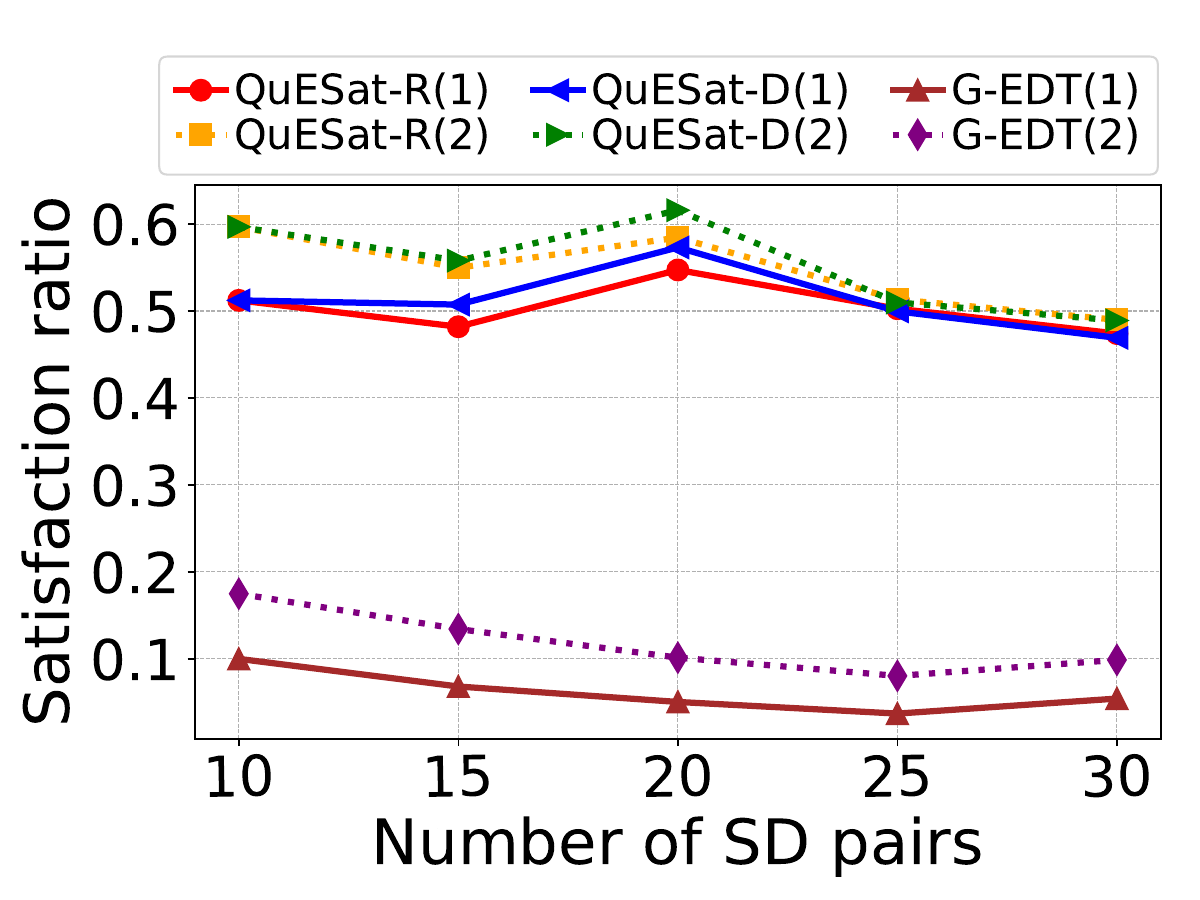}
\label{fig:pairs:satisfication_ratio}}
\vspace{-2mm}
\caption{Impact of number of SD pairs for QuESat and G-EDT}
\label{fig:pairs}
\end{figure}
\subsubsection{Comparing average throughput over time}
Fig.~\ref{fig:time_throughput}\subref{fig:time:average_throughput_commodities}--\subref{fig:time:average_throughput_fullgraph} show the average throughput over time in Scenario (1) and (2) respectively.
Each algorithm has a very short bootstrap period when the throughput increases because of the probabilistic protocol we implement following~\cite{Dai2020a, fendi}.
QuESat achieves multi-fold improvement over G-EDT in both scenarios, showing the significant advantage of utilizing direct satellite lightpaths with low loss for quantum communication.
The randomized and deterministic rounding algorithms achieve similar throughput hence either one can be used in practice.
Meanwhile, with more ground stations acting as repeaters in Scenario (2), the achieved throughput is higher than utilizing direct satellite lightpaths and fiber of SD pairs.
This is because more repeaters result in more flexible (re-)distribution of entanglements among SD pairs to maximize overall throughput.
\subsubsection{Comparing satisfaction ratio over time}
Fig.~\ref{fig:time_satis_ratio} shows similar results for the satisfaction ratio in the two scenarios as shown in Fig.~\ref{fig:time_throughput}.
QuESat significantly outperforms G-EDT using only ground fibers and involving more repeaters in Fig.~\ref{fig:time_satis_ratio}\subref{fig:time:satisfication_ratio_fullgraph} achieves better satisfaction ratio than Fig.~\ref{fig:time_satis_ratio}\subref{fig:time:satisfication_ratio_commodities}.

\subsubsection{Impact of distance}
To investigate the impact of distances on the ground repeater network, we scale the ground distances by a certain ratio to investigate when a ground repeater network would perform comparable to or better than the satellite network.
Fig.~\ref{fig:distance}\subref{fig:distance:averagethroughput} and Fig.~\ref{fig:distance}\subref{fig:distance:satisfication_ratio} show the average throughput and satisfaction ratio with varying distance ratios. 
Both average throughput and the satisfaction ratio decreased with increasing distances.
Even at relatively short distances such as the ratio of $10^{-2}$, QuESat still offers substantial advantage over using the ground network alone.
The results are due to two reasons: (i) As the distance ratio increases, the fiber links incur an exponentially larger loss, leading to a degraded possibility of successful ebit generation on the ground. 
(ii) Compared to fiber, entanglement distribution over satellite lightpaths is not easily affected by distance but mainly by the sequence of lenses they pass through. 
Overall, QuESat is not only crucial for long-distance entanglement distribution when the ground network has almost zero success but also substantially improves performance over relatively short distances, such as between nearby cities or within a large metropolitan area.

\subsubsection{Impact of number of SD pairs}
Fig.~\ref{fig:pairs}\subref{fig:pairs:averagethroughput} and Fig.~\ref{fig:pairs}\subref{fig:pairs:satisfication_ratio} show the average throughput and satisfaction ratio with varying number of SD pairs.
The average throughput increased with the increasing number of SD pairs in Fig.~\ref{fig:pairs}\subref{fig:pairs:averagethroughput}, while the satisfaction ratio fluctuated in Fig.~\ref{fig:pairs}\subref{fig:pairs:satisfication_ratio}. 
As the number of SD pairs increases, the total demand of SD pairs increases, and hence more lightpaths may be established, leading to higher throughput in QuESat.
With possibly more lightpaths, the satisfaction ratio remains generally stable until when all lightpaths deplete for QuESat, while G-EDT incurs a degraded satisfaction ratio due to the saturation of ground resources.

To summarize, we make three key observations from the evaluation:
(i) With the help of inter-satellites quantum transmission, \textbf{QuESat can satisfy significantly more entanglement demands} of SD pairs than the optimal throughput using ground fiber alone;
(ii) \textbf{Satellite lightpaths offer a flexible way to drastically improve entanglement throughput} over the ground network, even under frequent orbital dynamics and demand changes;
(iii) Considering different loss factors in space versus on the ground, QuESat achieves the most substantial improvement when \textbf{distributing long-distance, global-scale entanglements}, while the fiber-based architecture is more suitable and cost-friendly at short distances.



\section{Conclusions}
\label{sec:conclusions}
\noindent 
In this paper, we investigated a new hybrid ground-satellite quantum network architecture (QuESat) for entanglement distribution, integrating an on-the-ground fiber network with a global-scale passive optical network built with LEO satellites.
To account for satellite orbital dynamics and entanglement demand fluctuation, we formulated the lightpath provisioning and the entanglement distribution problems to enable efficient, global-scale entanglement distribution in the hybrid architecture.
We then developed a two-stage algorithm to dynamically configure the beam guides and distribute entanglements between ground stations.
A ground-satellite quantum network simulator was developed to simulate orbital dynamics, user demands and entanglement distribution both in space and on the ground.
Extensive simulation results showed the superior performance of our hybrid ground-satellite architecture in terms of average throughput and satisfaction ratio compared to ground repeater networks solely based on ground fiber.


\bibliographystyle{myIEEEtranS}
\clearpage


\begin{thebibliography}{10}
\providecommand{\url}[1]{#1}
\csname url@samestyle\endcsname
\providecommand{\newblock}{\relax}
\providecommand{\bibinfo}[2]{#2}
\providecommand{\BIBentrySTDinterwordspacing}{\spaceskip=0pt\relax}
\providecommand{\BIBentryALTinterwordstretchfactor}{4}
\providecommand{\BIBentryALTinterwordspacing}{\spaceskip=\fontdimen2\font plus
\BIBentryALTinterwordstretchfactor\fontdimen3\font minus \fontdimen4\font\relax}
\providecommand{\BIBforeignlanguage}[2]{{%
\expandafter\ifx\csname l@#1\endcsname\relax
\typeout{** WARNING: IEEEtranS.bst: No hyphenation pattern has been}%
\typeout{** loaded for the language `#1'. Using the pattern for}%
\typeout{** the default language instead.}%
\else
\language=\csname l@#1\endcsname
\fi
#2}}
\providecommand{\BIBdecl}{\relax}
\BIBdecl

\bibitem{gurobi}
\BIBentryALTinterwordspacing
``{Gurobi Optimizer},'' accessed 2024-04-10. URL: \url{http://www.gurobi.com/products/gurobi-optimizer}
\BIBentrySTDinterwordspacing

\bibitem{nv_1_46s}
M.~H. Abobeih, J.~Cramer, M.~A. Bakker, N.~Kalb, M.~Markham, D.~J. Twitchen, and T.~H. Taminiau, ``One-second coherence for a single electron spin coupled to a multi-qubit nuclear-spin environment,'' \emph{Nature Communications}, vol.~9, no.~1, p. 2552, 2018.

\bibitem{alicherry2012network}
M.~Alicherry and T.~Lakshman, ``Network aware resource allocation in distributed clouds,'' in \emph{IEEE INFOCOM}, 2012, pp. 963--971.

\bibitem{bayerbach2023bell}
M.~J. Bayerbach, S.~E. D’Aurelio, P.~van Loock, and S.~Barz, ``Bell-state measurement exceeding 50\% success probability with linear optics,'' \emph{Science Advances}, vol.~9, no.~32, 2023.

\bibitem{bennett2020quantum}
C.~H. Bennett and G.~Brassard, ``Quantum cryptography: Public key distribution and coin tossing,'' \emph{Theoretical Computer Science}, 2014.

\bibitem{bhattacherjee2019network}
D.~Bhattacherjee and A.~Singla, ``Network topology design at 27,000 km/hour,'' in \emph{ACM CoNEXT}, 2019, pp. 341--354.

\bibitem{nv_10}
C.~E. Bradley, J.~Randall, M.~H. Abobeih, R.~C. Berrevoets, M.~J. Degen, M.~A. Bakker, M.~Markham, D.~J. Twitchen, and T.~H. Taminiau, ``A ten-qubit solid-state spin register with quantum memory up to one minute,'' \emph{Physical Review X}, vol.~9, no.~3, p. 031045, 2019.

\bibitem{cacciapuoti2019quantum}
A.~S. Cacciapuoti, M.~Caleffi, F.~Tafuri, F.~S. Cataliotti, S.~Gherardini, and G.~Bianchi, ``{Quantum Internet: Networking challenges in distributed quantum computing},'' \emph{IEEE Network}, vol.~34, no.~1, pp. 137--143, 2019.

\bibitem{calsamiglia2001maximum}
J.~Calsamiglia and N.~L{\"u}tkenhaus, ``Maximum efficiency of a linear-optical bell-state analyzer,'' \emph{Applied Physics B}, vol.~72, pp. 67--71, 2001.

\bibitem{chang2023entanglement_sat}
A.~Chang, Y.~Wan, G.~Xue, and A.~Sen, ``Entanglement distribution in satellite-based dynamic quantum networks,'' \emph{IEEE Network}, 2023.

\bibitem{chen2022ddka}
L.~Chen, Q.~Chen, M.~Zhao, J.~Chen, S.~Liu, and Y.~Zhao, ``{DDKA-QKDN}: Dynamic on-demand key allocation scheme for quantum internet of things secured by qkd network,'' \emph{Entropy}, vol.~24, no.~2, p. 149, 2022.

\bibitem{chicago-quantum-MAN}
J.~Chung, E.~M. Eastman, G.~S. Kanter, K.~Kapoor, N.~Lauk, C.~H. Pena, R.~K. Plunkett, N.~Sinclair, J.~M. Thomas, R.~Valivarthi \emph{et~al.}, ``Design and implementation of the illinois express quantum metropolitan area network,'' \emph{IEEE Transactions on Quantum Engineering}, 2022.

\bibitem{cicconetti2022resource}
C.~Cicconetti, M.~Conti, and A.~Passarella, ``Resource allocation in quantum networks for distributed quantum computing,'' \emph{arXiv preprint arXiv:2203.05844}, 2022.

\bibitem{dahlberg2019link}
A.~Dahlberg, M.~Skrzypczyk, T.~Coopmans, L.~Wubben, F.~Rozp{e}dek, M.~Pompili, A.~Stolk, P.~Pawe{\l}czak, R.~Knegjens, J.~de~Oliveira~Filho \emph{et~al.}, ``A link layer protocol for quantum networks,'' in \emph{ACM SIGCOMM}, 2019, pp. 159--173.

\bibitem{Dai2020a}
W.~Dai, T.~Peng, and M.~Z. Win, ``Optimal protocols for remote entanglement distribution,'' in \emph{IEEE ICNC}, 2020, pp. 1014--1019.

\bibitem{quantum-queueing-delay}
------, ``Quantum queuing delay,'' \emph{IEEE Journal on Selected Areas in Communications}, vol.~38, no.~3, pp. 605--618, 2020.

\bibitem{dubhashi1998concentration}
D.~Dubhashi and A.~Panconesi, ``Concentration of measure for the analysis of randomised algorithms,'' \emph{Draft Manuscript}, 1998.

\bibitem{Dur1999}
W.~Dur, H.-J. Briegel, J.~I. Cirac, and P.~Zoller, ``Quantum repeaters based on entanglement purification,'' \emph{Physical Review A}, 1999.

\bibitem{goswami2023satellite}
S.~Goswami and S.~Dhara, ``Satellite-relayed global quantum communication without quantum memory,'' \emph{Physical Review A}, 2023.

\bibitem{fendi}
H.~Gu, Z.~Li, R.~Yu, X.~Wang, F.~Zhou, J.~Liu, and G.~Xue, ``Fendi: Toward high-fidelity entanglement distribution in the quantum internet,'' \emph{IEEE/ACM Transactions on Networking}, 2024.

\bibitem{hein2005entanglement}
M.~Hein, W.~D{\"u}r, and H.-J. Briegel, ``Entanglement properties of multipartite entangled states under the influence of decoherence,'' \emph{Physical Review A}, vol.~71, no.~3, p. 032350, 2005.

\bibitem{huang2024vacuum}
Y.~Huang, F.~Salces-Carcoba, R.~X. Adhikari, A.~H. Safavi-Naeini, and L.~Jiang, ``Vacuum beam guide for large scale quantum networks,'' \emph{Physical Review Letters}, vol. 133, no.~2, p. 020801, 2024.

\bibitem{kar2003routing}
K.~Kar, M.~Kodialam, and T.~Lakshman, ``Routing restorable bandwidth guaranteed connections using maximum 2-route flows,'' \emph{IEEE/ACM Transactions on Networking}, vol.~11, no.~5, pp. 772--781, 2003.

\bibitem{satellite_simulator}
B.~S. Kempton, ``A simulation tool to study routing in large broadband satellite networks,'' 2020.

\bibitem{kodialam2003dynamic}
M.~Kodialam and T.~Lakshman, ``Dynamic routing of restorable bandwidth-guaranteed tunnels using aggregated network resource usage information,'' \emph{IEEE/ACM Transactions on Networking}, 2003.

\bibitem{spdc}
P.~G. Kwiat, K.~Mattle, H.~Weinfurter, A.~Zeilinger, A.~V. Sergienko, and Y.~Shih, ``New high-intensity source of polarization-entangled photon pairs,'' \emph{Physical Review Letters}, vol.~75, no.~24, p. 4337, 1995.

\bibitem{lai2023achieving}
Z.~Lai, H.~Li, Y.~Wang, Q.~Wu, Y.~Deng, J.~Liu, Y.~Li, and J.~Wu, ``Achieving resilient and performance-guaranteed routing in space-terrestrial integrated networks,'' in \emph{IEEE INFOCOM}, 2023, pp. 1--10.

\bibitem{four-wave}
X.~Li, P.~L. Voss, J.~E. Sharping, and P.~Kumar, ``Optical-fiber source of polarization-entangled photons in the 1550 nm telecom band,'' \emph{Physical Review Letters}, vol.~94, no.~5, p. 053601, 2005.

\bibitem{li2021internet}
Y.~Li, H.~Li, L.~Liu, W.~Liu, J.~Liu, J.~Wu, Q.~Wu, J.~Liu, and Z.~Lai, ``Internet in space for terrestrial users via cyber-physical convergence,'' in \emph{ACM HotNets}, 2021, pp. 163--170.

\bibitem{liu2021optical}
H.-Y. Liu, X.-H. Tian, C.~Gu, P.~Fan, X.~Ni, R.~Yang, J.-N. Zhang, M.~Hu, J.~Guo, X.~Cao \emph{et~al.}, ``Optical-relayed entanglement distribution using drones as mobile nodes,'' \emph{Physical Review Letters}, 2021.

\bibitem{liu2024quantum}
M.~Liu, Z.~Li, K.~Cai, J.~Allcock, S.~Zhang, and J.~Lui, ``Quantum bgp with online path selection via network benchmarking,'' in \emph{IEEE INFOCOM}, 2024.

\bibitem{liu2019constant}
M.~Liu, A.~W. Richa, M.~Rost, and S.~Schmid, ``A constant approximation for maximum throughput multicommodity routing and its application to delay-tolerant network scheduling,'' in \emph{IEEE INFOCOM}, 2019.

\bibitem{mao2023qubit}
Y.~Mao, Y.~Liu, and Y.~Yang, ``Qubit allocation for distributed quantum computing,'' in \emph{IEEE INFOCOM}, 2023, pp. 1--10.

\bibitem{starlink}
J.~C. McDowell, ``The low earth orbit satellite population and impacts of the spacex starlink constellation,'' \emph{The Astrophysical Journal Letters}, vol. 892, no.~2, p. L36, 2020.

\bibitem{nitish_satellite_2022}
N.~K. Panigrahy, P.~Dhara, D.~Towsley, S.~Guha, and L.~Tassiulas, ``Optimal entanglement distribution using satellite based quantum networks,'' in \emph{IEEE INFOCOM WKSHPS}, 2022, pp. 1--6.

\bibitem{panigrahy2022capacity}
N.~K. Panigrahy, T.~Vasantam, D.~Towsley, and L.~Tassiulas, ``On the capacity region of a quantum switch with entanglement purification,'' in \emph{IEEE INFOCOM}, 2023.

\bibitem{pant2019routing}
M.~Pant, H.~Krovi, D.~Towsley, L.~Tassiulas, L.~Jiang, P.~Basu, D.~Englund, and S.~Guha, ``Routing entanglement in the quantum internet,'' \emph{npj Quantum Information}, vol.~5, no.~1, pp. 1--9, 2019.

\bibitem{peev2009secoqc}
M.~Peev, C.~Pacher, R.~All{\'e}aume, C.~Barreiro, J.~Bouda, W.~Boxleitner, T.~Debuisschert, E.~Diamanti, M.~Dianati, J.~Dynes \emph{et~al.}, ``The secoqc quantum key distribution network in vienna,'' \emph{New Journal of Physics}, vol.~11, no.~7, p. 075001, 2009.

\bibitem{pirandola2019end}
S.~Pirandola, ``End-to-end capacities of a quantum communication network,'' \emph{Communications Physics}, vol.~2, no.~1, pp. 1--10, 2019.

\bibitem{photon-loss}
S.~Pirandola, R.~Laurenza, C.~Ottaviani, and L.~Banchi, ``Fundamental limits of repeaterless quantum communications,'' \emph{Nature Communications}, vol.~8, no.~1, p. 15043, 2017.

\bibitem{pouryousef2022quantum}
S.~Pouryousef, N.~K. Panigrahy, and D.~Towsley, ``A quantum overlay network for efficient entanglement distribution,'' \emph{IEEE INFOCOM}, 2023.

\bibitem{raghavan1988probabilistic}
P.~Raghavan, ``Probabilistic construction of deterministic algorithms: approximating packing integer programs,'' \emph{Journal of Computer and System Sciences}, vol.~37, no.~2, pp. 130--143, 1988.

\bibitem{sasaki2011field}
M.~Sasaki, M.~Fujiwara, H.~Ishizuka, W.~Klaus, K.~Wakui, M.~Takeoka, S.~Miki, T.~Yamashita, Z.~Wang, A.~Tanaka \emph{et~al.}, ``{Field test of quantum key distribution in the Tokyo QKD Network},'' \emph{Optics Express}, 2011.

\bibitem{mip}
A.~Schrijver, \emph{Theory of linear and integer programming}.\hskip 1em plus 0.5em minus 0.4em\relax Wiley, 1998.

\bibitem{sen2023quantum_satellite}
A.~Sen, C.~Sumnicht, S.~Choudhuri, A.~Chang, and G.~Xue, ``Quantum communication in 6g satellite networks: Entanglement distribution across changing topologies,'' \emph{IEEE ICC}, 2024.

\bibitem{shi2020concurrent}
S.~Shi and C.~Qian, ``Concurrent entanglement routing for quantum networks: Model and designs,'' in \emph{ACM SIGCOMM}, 2020, pp. 62--75.

\bibitem{tune2015spatiotemporal}
P.~Tune and M.~Roughan, ``Spatiotemporal traffic matrix synthesis,'' in \emph{ACM SIGCOMM}, 2015, pp. 579--592.

\bibitem{zhao2024Routing}
Y.~Wang, Y.~Zhao, L.~Huang, and C.~Qiao, ``{Routing and Wavelength Assignment for Entanglement Swapping of Photonic Qubits},'' in \emph{IEEE INFOCOM}, 2024.

\bibitem{QCNC_sat}
X.~Wei, L.~Fan, Y.~Guo, Z.~Han, and Y.~Wang, ``Optimizing satellite-based entanglement distribution in quantum networks via quantum-assisted approaches,'' \emph{IEEE QCNC}, 2024.

\bibitem{wei2024optimal_sat}
X.~Wei, J.~Liu, L.~Fan, Y.~Guo, Z.~Han, and Y.~Wang, ``Optimal entanglement distribution problem in satellite-based quantum networks,'' \emph{IEEE Network}, 2024.

\bibitem{zhao2023scheduling}
L.~Yang, Y.~Zhao, L.~Huang, and C.~Qiao, ``Asynchronous entanglement provisioning and routing for distributed quantum computing,'' in \emph{IEEE INFOCOM}, 2023.

\bibitem{yin2017satellite}
J.~Yin, Y.~Cao, Y.-H. Li, S.-K. Liao, L.~Zhang, J.-G. Ren, W.-Q. Cai, W.-Y. Liu, B.~Li, H.~Dai \emph{et~al.}, ``Satellite-based entanglement distribution over 1200 kilometers,'' \emph{Science}, vol. 356, no. 6343, pp. 1140--1144, 2017.

\bibitem{yin2017satellite_2}
J.~Yin, Y.~Cao, Y.-H. Li, J.-G. Ren, S.-K. Liao, L.~Zhang, W.-Q. Cai, W.-Y. Liu, B.~Li, H.~Dai \emph{et~al.}, ``Satellite-to-ground entanglement-based quantum key distribution,'' \emph{Physical Review Letters}, 2017.

\bibitem{zeng2022multi}
Y.~Zeng, J.~Zhang, J.~Liu, Z.~Liu, and Y.~Yang, ``Multi-entanglement routing design over quantum networks,'' in \emph{IEEE INFOCOM}, 2022.

\end{thebibliography}



\end{document}